\documentclass{article}

\usepackage[numbers]{natbib}         
\usepackage[colorlinks]{hyperref}    
\usepackage[english]{babel} 
\usepackage{amssymb}
\usepackage{amsmath}
\usepackage{txfonts}
\usepackage{mathdots}
\usepackage[classicReIm]{kpfonts}
\usepackage[dvips]{graphicx} 
\usepackage[a4paper, portrait, margin=1in]{geometry}
\usepackage{tabularx}
\usepackage{siunitx}
\usepackage{longtable}
\usepackage{multirow}
\usepackage{makecell}
\begin{document}


\textbf{\Large Medical Imaging Synthesis using Deep Learning and its Clinical Applications: A Review}

\textbf{ }

Tonghe Wang${}^{1,2}$, Yang Lei${}^{1}$, Yabo Fu${}^{1}$, Walter J. Curran${}^{1,2}$, Tian Liu${}^{1,2}$ and Xiaofeng Yang${}^{1,2}$*

${}^{1}$Department of Radiation Oncology, Emory University, Atlanta, GA 

${}^{2}$Winship Cancer Institute, Emory University, Atlanta, GA

\noindent 
\bigbreak
\bigbreak
\bigbreak

\textbf{*Corresponding author: }

Xiaofeng Yang, PhD

Department of Radiation Oncology

Emory University School of Medicine

1365 Clifton Road NE

Atlanta, GA 30322

E-mail: xiaofeng.yang@emory.edu

\bigbreak
\bigbreak
\bigbreak
\bigbreak
\bigbreak
\bigbreak

\textbf{Abstract}

This paper reviewed the deep learning-based studies for medical imaging synthesis and its clinical application. Specifically, we summarized the recent developments of deep learning-based methods in inter- and intra-modality image synthesis by listing and highlighting the proposed methods, study designs and reported performances with related clinical applications on representative studies. The challenges among the reviewed studies were summarized in the discussion part.

\noindent \eject 

\noindent 
\section{ INTRODUCTION}

Image synthesis between different medical imaging modalities/protocols is an active research field with great clinical interest in radiation oncology and radiology. It aims to facilitate a specific clinical workflow by bypassing or replacing a certain imaging procedure when the acquisition is infeasible, costs additional time/labor/expense, has ionizing radiation exposure, or introduces uncertainty from image registration between different modalities. The proposed benefit has raised increasing interest in a number of potential clinical applications such as magnetic resonance imaging (MRI)-only radiation therapy treatment planning, positron emission tomography (PET)/MRI scanning, and etc. 

 \bigbreak
       Image synthesis with its potential applications has been investigated for decades. The conventional methods usually rely on models with explicit human-defined rules about the conversion of images from one modality to the other. These models are usually application-specific depending on the unique characteristics of the involved pair of imaging modalities, thus can be diverse in methodologies and complexities. It is also hard to build such a model when the two imaging modalities provide distinct information, such as anatomic imaging and functional imaging. This is partially why the majority of these studies are limited to image synthesis between computed tomography (CT) images from MRI.\cite{RN2311} These methods usually require case-by-case parameters tuning for optimal performance.
 \bigbreak
       Owing to the widespread success of machine learning in computer vision field in recent years, the latest breakthrough in artificial intelligence has been integrated into medical image synthesis. In addition to CT-MRI synthesis, image synthesis in other imaging modalities such as PET and cone-beam CT (CBCT) is now viable. As a result, more and more applications could benefit from the recent advancements of image synthesis techniques.\cite{RN5361, RN1679, RN5342} Deep learning, as a large subset of machine learning and artificial intelligence, is dominating in this field in the past several years. Deep learning utilizes neural network with many layers containing huge number of neurons to extract useful features from images. Various networks and architectures have been proposed for better performance on different tasks. Deep learning-based image synthesis methods usually share a common framework that uses a data-driven approach for image intensity mapping. The workflow usually consists of a training stage for the network to learn the mapping between the input and its target, and a predication stage to synthesize the target from an input. Compared with conventional model-based methods, deep learning-based methods are more generalizable since the same network and architecture for a pair of image modalities can be applied to different pairs of image modalities with minimal adjustment. This allows rapid expansion of applications using a similar methodology to a variety of imaging modalities that are clinically desired for image synthesis. The performance of the deep learning-based methods largely depends on the representativeness of the training datasets rather than case-specific parameters. Although the network training may require lots of efforts in collecting and cleaning training datasets, the prediction usually takes only a few seconds. Due to these advantages, deep learning-based methods have attracted great research and clinical interest in medical imaging and radiation therapy. 
 \bigbreak
       In this paper, we systematically reviewed the emerging deep learning-based methods and applications for medical image synthesis. Specifically, we categorized the recent literatures based on their deep learning properties and highlighted their contributions. The clinical scenario of applications was then summarized with challenges and concerns identified. The trend and future direction on this topic were discussed in the end of this review. 

\bigbreak
\noindent 
\section{ LITERATURE SEARCHING}

We defined the scope of this review study to include both inter-modality and intra-modality image synthesis using deep learning method. Inter-modality applications included studies about the image synthesis between two different imaging modalities. Intra-modality applications included studies that transform images between two different protocols of a same imaging modality, such as between different MR imaging sequences, or the restoration of images from low quality protocol to high quality protocol. Studies solely aiming for image quality improvement such as image denoising and artifact correction were not included in this study. Conference abstracts and proceedings were not considered due to the lack of strict peer review process in study design and reported results.   
 \bigbreak
       Peer-reviewed journal publications were searched on PubMed using the criteria in title or abstract as of February 2020 : (“pseudo” OR “synth*” OR “reconstruct*” OR “transform” OR “restor*” OR “correct*” OR “generat*”) AND “deep” AND “learning” AND (“CT” OR “MR” OR “MRI” OR “PET” OR “SPECT” OR “Ultrasound”). The search yielded 681 records. We manually screened each record and removed ineligible ones, and the remaining 70 articles were included in this review study. We also performed a citation search on the identified literatures, and additional 41 articles were included. Therefore, 111 articles in total were included in this review. Compared with current review papers on this topic,\cite{RN4605} this review study is comprehensive in covering more articles by a systematic review approach. Figure 1 shows the number of reviewed articles in each year. With the earliest one published in 2017 and an increment of about 25 per year, the number of publications on this topic has increased linearly. The number of articles in the first two months of 2020 has surpassed the total number in 2017. 

\bigbreak
\begin{figure}
\centering
\noindent \includegraphics*[width=5.50in,  keepaspectratio=true]{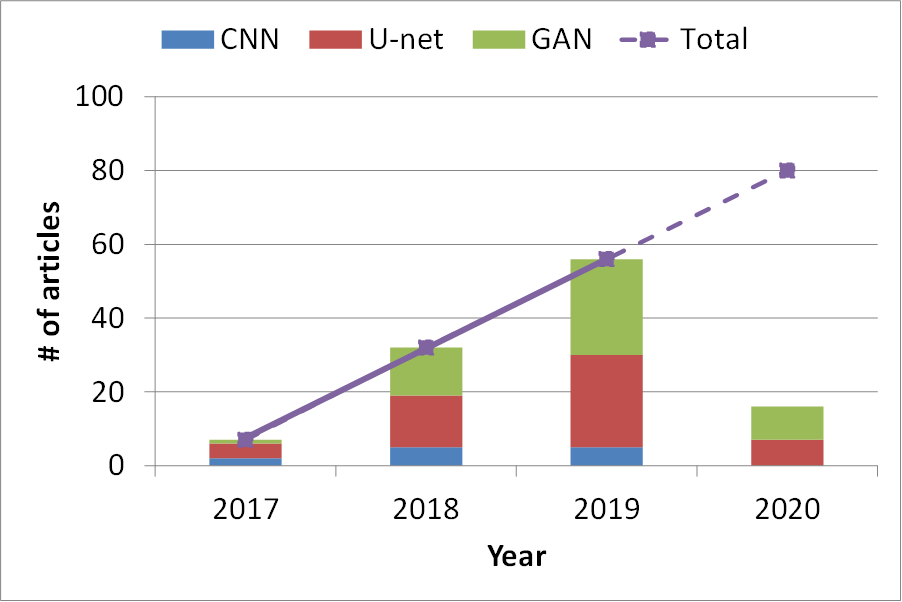}

\noindent Figure 1. Number of peer-reviewed articles in medical imaging synthesis using deep learning with different neural networks. This study only covers the first two month of 2020. The dashed line predicting the total number of articles in 2020 is a linear extrapolation based on previous years. 
\end{figure}

\bigbreak
\noindent 
\section{ DEEP LEARNING METHODS}

The framework of the reviewed studies can be grouped into three categories: Convolutional Neural Network (CNN), U-net, and Generative Adversarial Network (GAN). The pie chart of the three groups shown in Figure 2 indicates that U-net and GAN studies, which are close in total numbers, are the mainstream that accounts for about 90\%. Figure 1 also demonstrates that the studies using U-net and GAN keep increasing since 2017, with GAN in a larger rate than U-net. Most methods of all these three categories are supervised learning. Three out of 111 studies used an unsupervised strategy that learned image translation from unpaired datasets. The three categories, CNN, U-net and GAN, have an increasing complexity and are not completely distinct from each other. A review of methods in each category is provided in this section.  

\bigbreak
\begin{figure}
\centering
\noindent \includegraphics*[width=5.50in, keepaspectratio=true]{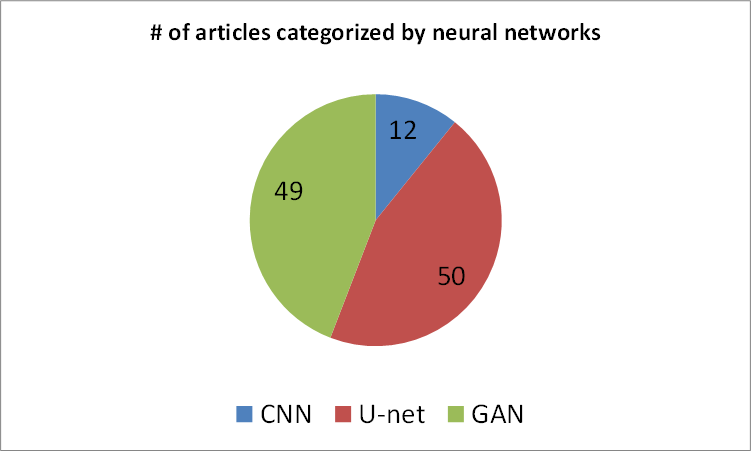}

\noindent Figure 2. Pie chart of numbers of articles in different categories of neural networks. 
\end{figure}

\bigbreak
\noindent 
\subsection{Convolutional Neural Network}

Convolutional Neural Network (CNN) is a class of deep neutral networks that use convolution kernels to explore the spatially local image patterns. It consists of an input, an output, and multiple hidden layers. The hidden layers contain a series of convolutional layers that convolve the input with trainable convolution kernels and pass the feature maps to the next layer. Rectified linear unit (ReLU) layer is the commonly used activation layer due to its computational simplicity, representational sparsity and linearity. Dropout layers are commonly used to reduce chances of overfitting. Batch normalization layer is usually employed to reduce internal covariate shift of the training datasets for improved robustness and faster convergency. To save memory, the large size of image is typically reduced after pooling and convolution layers to allow larger number of feature maps and deeper networks. With multiple hidden convolutional layers, a hierarchy of increasingly complex features with high-level abstraction is then extracted. During the training process, iterative adjustments are made on the weights and biases of the kernels of these convolutional layers until the loss function is minimized. These weights and biases are called trainable parameters of networks. Gradient descent methods, such as Adam optimizer, are used to update trainable parameters of our networks. A simple basic CNN is composed of several connected convolutional layers to map input to output. Nonetheless, very few studies directly employ CNN in its basic form. Most of the reviewed studies used variants of basic CNN for better performance. For example, ResNet was chosen in a few studies since it has short-cut connections for each block that skip one or more layers, which eases the training of the deep network without adding extra parameters or computational complexity.\cite{RN5204, RN4731, RN4612} It also allows feature maps from the initial layers that usually contain fine details to be easily propagated to the deeper layers. CNN and its variants are commonly utilized as a basic component in advanced architectures such as the ones listed in the following. 

\bigbreak
\noindent 
\subsection{U-net}

As one of the first several studies employing deep learning in image synthesis, Han used CNN in synthesizing CT from MR images by adopting and modifying a U-net architecture.\cite{RN5023} The U-net model used in the study of Han has an encoding and a decoding part. The encoding part acts as the CNN mentioned above that extract hierarchical features from an input MR image with convolutional, batch normalization, ReLU and pooling layers, and the decoding part, a mirrored version of the encoding part that replaces pooling layers with deconvolution layers, transforms the features and reconstructs the predicted CT images from low to high resolution levels. The two parts are connected through short-cuts on multiple layers such that high-resolution features from encoding part can be used as extra inputs in the decoding part. Moreover, the model removed fully-connected layers such that the number of parameters was highly reduced. In their study, the model was trained by pairs of MR and CT 2D slices. During the learning process, a loss function of mean absolute error (MAE) between prediction and ground truth was minimized. The usage of this L1-norm as loss function can help the learning be more robust to noise, artifacts and misalignment among the training images. 
\bigbreak
       Most of the studies applying U-net into their model generally followed the above architecture, with many variants and improvements proposed and studied. For example, compared with the model of Han, Liu \textit{et al.} applied a similar encoder and decoder model while without the skip connection.\cite{RN5224, RN5002} In their MR-based sCT study, instead of using CT images as training and prediction target, they used discretized maps from CTs by labeling three materials, which converted the CT synthesis into a segmentation problem. Thus, a multi-class soft-max classifier was added on the final layer of the decoder to produce class probabilities for each voxel. Another difference in Ref. \cite{RN5224} is that since this model was based on 2D image slices, a fully connected conditional random field was added to consider the 3D contextual relationship between voxels since it can take the output of the model and take the original 3D volume to build the pairwise potentials on all pairs of voxels. Dong \textit{et al.} found that the information that the long skip connection in U-net concatenates from the encoding path is high frequency, which often includes irrelevant components from noisy input images. In order to address this issue, they used a self-attention strategy that uses the feature maps extracted from coarse scale to eliminate noisy responses prior to the concatenation. This self-attention U-net is able to highlight the most salient features from the coding path.\cite{RN4754} Hwang \textit{et al.} also noticed the noise propagation from high frequency feature, thus they only used the contracting path in deeper layers.\cite{RN5256} 
\bigbreak
       The choice of building blocks has also been investigated. Fu \textit{et al.} made a few improvements based on the architecture of Han. For example, batch normalization layers were replaced with instance normalization layers for a better performance when trained with small batch size. The unpooling layers, which produce sparse feature maps, were also replaced with deconvolutional layers that produce dense feature maps. The skip connections were replaced with residual short-cuts, which was inspired by ResNet, to further save computational memory. \cite{RN4879} The ReLU layer was also replaced to be a generalized parametric ReLU (PReLU) in the study of Neppl \textit{et al.} to adaptively adjust the activation function.\cite{RN4865} Torrado-Carvajal \textit{et al.} added a dropout layer before the first transposed convolution of the decoder to avoid overfitting.\cite{RN5172} 
\bigbreak
       Various loss functions have been investigated in the reviewed studies. In addition to the most commonly used L1-norm and L2-norm functions that enforce voxel-wise similarity, other functions that describe different image properties are usually combined into the total loss function. For example, Leynes \textit{et al.} used a total loss function which was a sum of MAE loss, gradient difference loss and Laplacian difference loss, the last two of which help improve image sharpness.\cite{RN4993} Similarly, Chen \textit{et al.} combines the MAE loss with structure dissimilarity loss to encourage whole-structure-wise similarity.\cite{RN4682} L2 regularization has also been incorporated into the loss function in a few studies to avoid overfitting.\cite{RN4722, RN5173} Kazemifar \textit{et al.} used mutual information, which has been widely used in image registration, in their loss function, and demonstrated its advantages over MAE loss in better compensating the misalignment between CT and MR images. Largent \textit{et al.} introduced a perceptual loss which can mimic human visual perception using similar features rather than only intensities, into their U-net. The perceptual loss was proposed to implement in three different ways with increasing complexity: on a single convolutional layer, on multiple layers with uniform weights, and on multiple layers with different weights that give more importance to the layers yielding the lower MAE.\cite{RN4787}

\bigbreak
\noindent 
\subsection{Generative Adversarial Network}

Generative Adversarial Network (GAN) is composed of a generative network and a discriminative network that are trained simultaneously. The generative network is trained to generate synthetic images, and the discriminative network is trained to classify an input image as real or synthetic. The training goal of GAN is to let the generative network produce synthetic images as realistic as possible to fool the discriminator, while let the discriminative network to distinguish the synthetic images from real images. In this way, blurry synthetic images can be easily identified by the discriminator since they look considerably fake. This conflict goal explains the name of “adversarial”. Both networks are trained better and better when they compete against each other until equilibrium is reached. In the prediction stage, the trained generative network is applied on new incoming image. 
\bigbreak
       Similar to CNN, GAN has also been used in one of the earliest publications in medical image synthesis using deep learning. Nie \textit{et al.} used a fully convolutional network (a variant of CNN) and a CNN for the generative and discriminative, respectively.\cite{RN5196} The loss function of the discriminative network was binary cross entropy, which was minimized between its decisions and correct labels. Similarly, in generative network, a binary cross entropy between the decision by discriminative network and the wrong label for the generated images, was added into the loss function. Since the network in this study was trained in a patch-to-patch manner that may limit the context information available in the training samples, an auto-context model was employed to refine the results. 
\bigbreak
       Based on the basic architecture of GAN, many variants have been designed and investigated. Emami \textit{et al.} adopted conditional GAN (cGAN) in CT synthesis from MR.\cite{RN5204} Unlike unconditional GAN, both the generative and discriminative networks of cGAN observe the input images (e.g. the MR images in CT synthesis from MR). It can be formulated by condition the loss function of discriminator on the input images, which has been proved to be more suitable for image-to-image translation tasks.\cite{RN5302} Liang et al adopted CycleGAN in their CBCT-based synthetic CT study.\cite{RN4915} The CycleGAN includes two generators which are CBCT-CT generator and CT-CBCT generator and two discriminators which are real CT-synthetic CT discriminator and real CBCT-synthetic CBCT discriminator. In the first cycle, the input CBCT is fed into the CBCT-CT generator to synthesize CT, and then the synthetic CT is fed into the CT-CBCT generator to generate cycle CBCT, which is supposed to be same as the input CBCT. The cycle CBCT is compared to the original input CBCT to generate CBCT cycle consistent loss. Meanwhile, the real CT-synthetic CT discriminator distinguishes between the real CT and the synthetic CT to generate CT adversarial loss. To encourage one-to-one mapping between CT and CBCT, a second cycle transformation from CT to CBCT is performed. The second cycle is same as the first cycle, except the roles of CBCT and CT are swapped, i.e. real CT is fed into the same CT-CBCT generator to synthesize CBCT, and then the synthetic CBCT is fed into the same CBCT-CT generator to generate cycle CT. The cycle CT is compared to the real CT to generate CT cycle consistent loss. The real CBCT-synthetic CBCT discriminator distinguishes between the CBCT and the synthetic CBCT to generate CBCT adversarial loss. Unlike GAN, the CycleGAN couples an inverse mapping network by introducing a cycle consistency loss which enhances the network performance, especially when paired training CT/CBCT images are absent. As a result, CycleGAN can tolerate certain level of misalignment in the paired training dataset. This property of CycleGAN is attractive to inter-modality synthesis since misalignment in the training datasets are sometimes inevitable due to the unavailability of exact matching image pairs. In many studies, training images are still paired by registration to preserve quantitative pixel values, remove large geometric mismatch to allow network to focus on mapping details and accelerate training.\cite{RN4884} 
\bigbreak
       Different structures of the feature extraction blocks were found to be useful for different applications. A group of studies showed that CNN with residual blocks can achieve promising results in image transforming tasks where source and target images are largely similar, such as between CT and CBCT, non-attenuation corrected (NAC) PET and attenuation corrected (AC) PET, and low-counting PET and full-counting PET. Since these pairs of images have similar image appearance but are different quantitatively, residual blocks were integrated into the network to learn the differences between the pairs; each residual block includes a residual connection and multiple hidden layers. An input bypasses these hidden layers through the residual connection, thus the hidden layers enforces learner to minimize a residual image between the source and target images, which usually is noise and artifacts.\cite{RN4668, RN4884, RN4770, RN5305, RN4740} In contrast, dense block concatenates outputs from previous layers instead of using the summation, to connect adjacent layers in a feed-forward fashion. It is able to capture multi-frequency (high frequency and low frequency) information to well-represent the mapping from source image modality to target image modality, thus it is commonly used in inter-modality image synthesis such as MR-to-CT, and PET-to-CT.\cite{RN4750, RN4754, RN4913, RN4888, RN4793, RN4904} 
\bigbreak
       CNN and its variants are commonly used for the generative and the discriminative networks. Emami \textit{et al.} used ResNet for its generative network.\cite{RN5204} They removed the fully connected layers and added two transposed convolutional layers after residue blocks as deconvolution. Kim \textit{et al.} combined the U-net and the residual training scheme in the generative network.\cite{RN5228} Olberg \textit{et al.} proposed a deep spatial pyramid convolutional framework that includes an atrous spatial pyramid pooling module in a U-net architecture. The module performs atrous convolution at multiple rates in parallel such that multiscale features can be exploited to characterize a single pixel.\cite{RN4842} The encoder is then able to capture rich multiscale contextual information, which aids the image translation. Compared with generator, the discriminator is mostly implemented in a simpler form. A typical example is a few down-sampling convolutional layers followed by a sigmoid layer to binarize the output, as proposed by Liu \textit{et al.}\cite{RN4793} 
\bigbreak
       In addition to the image quality and accuracy loss functions as in U-net, adversarial loss functions are incorporated in GAN and its variants. The adversarial term, unlike the reconstruction term that represents image intensity accuracy, reflects the correct or wrong decision that the discriminator makes on real image or synthetic image. Apart from the binary cross entropy mentioned above or a similar form of sigmoid cross entropy, the negative log likelihood functions from the original publication of GAN in computer vision are also widely used. However, the training process may suffer from divergence caused by vanishing gradient and mode collapse when discriminator is trained to be optimal for a fixed generator.\cite{RN5475} Emami \textit{et al.} proposed to use least square loss that has been shown to be more stable during training and generates higher quality results.\cite{RN5204} Loss function using Wasserstein distance can be another option since it has smoother gradient flow and better convergence than the original one.\cite{RN5475} It has also been shown that in GAN, simply providing the true or fake labels by the discriminator may not be sufficient for the generator to improve, which causes instability in training due to gradient vanishing and exploding. Ouyang \textit{et al.} employed a feature matching technique by specifying a new objective such that the generator encourages the synthesized images to match the expected value of features on the intermediate layers of discriminator, instead of directly maximizing the final output of the discriminator.\cite{RN4908} 

\bigbreak
\noindent 
\subsection{Other}

In addition to the above architectures, other designs have also been proposed to adapt to specific applications in the reviewed studies. For example, Zhang \textit{et al.} proposed a dual-domain CNN framework that uses two parallel CNNs in spatial and frequency domains respectively and interacts with each other by Fourier transform for generating synthetic 7T MRI from 3TMRI.\cite{RN4894}  The additional integration of frequency domain was proved to be superior to using spatial domain alone in synthesis accuracy. In the study of reconstructing ultra-low-dose amyloid PET reconstruction by Ouyang \textit{et al.}, a pretrained classifier that predicts the amyloid status (positive or negative) is incorporated into a GAN-based network. The pretrained amyloid status classifier acts as a feature extractor and provides feature maps in the calculation of perceptual loss combined in GAN. 
\bigbreak
       Using images from multiple modalities as input in deep learning network has been shown effective in providing more useful features for learning and testing in several studies. These multi-modality images are usually treated as inputs with multiple channels in the first layer, each of which has a spatial invariant kernel applied for convolution on the entire image. Wang \textit{et al.} claimed that the contributions of different modalities could vary at different locations, thus they added a locality adaptive fusion network that takes two modalities (a low counting PET and a T1-weighted MRI in their study) as input to generate a fused image by learning different convolutional kernels at different image locations. The fused image is then fed into the generative network in GAN architecture.\cite{RN4931} In contrast to common multi-channel inputs in a single path, Tie \textit{et al.} used three MR images with different contrast as multi-channel inputs in a multi-path architecture which has three training paths in the encoder and each channel has its own feature network.\cite{RN4612} The separate image feature extractions on different MR images are able to avoid the loss of unique features that may be merged in the low level. 

\bigbreak
\noindent 
\section{ APPLICATION AREAS}

The reviewed articles were categorized into two main groups in this study based on their study objectives: inter-modality (56\%) and intra-modality (44\%). In each group, there are subgroups that specify the imaging modalities and clinical applications.  

\bigbreak
\begin{figure}
\centering
\noindent \includegraphics*[width=6.50in,  keepaspectratio=true]{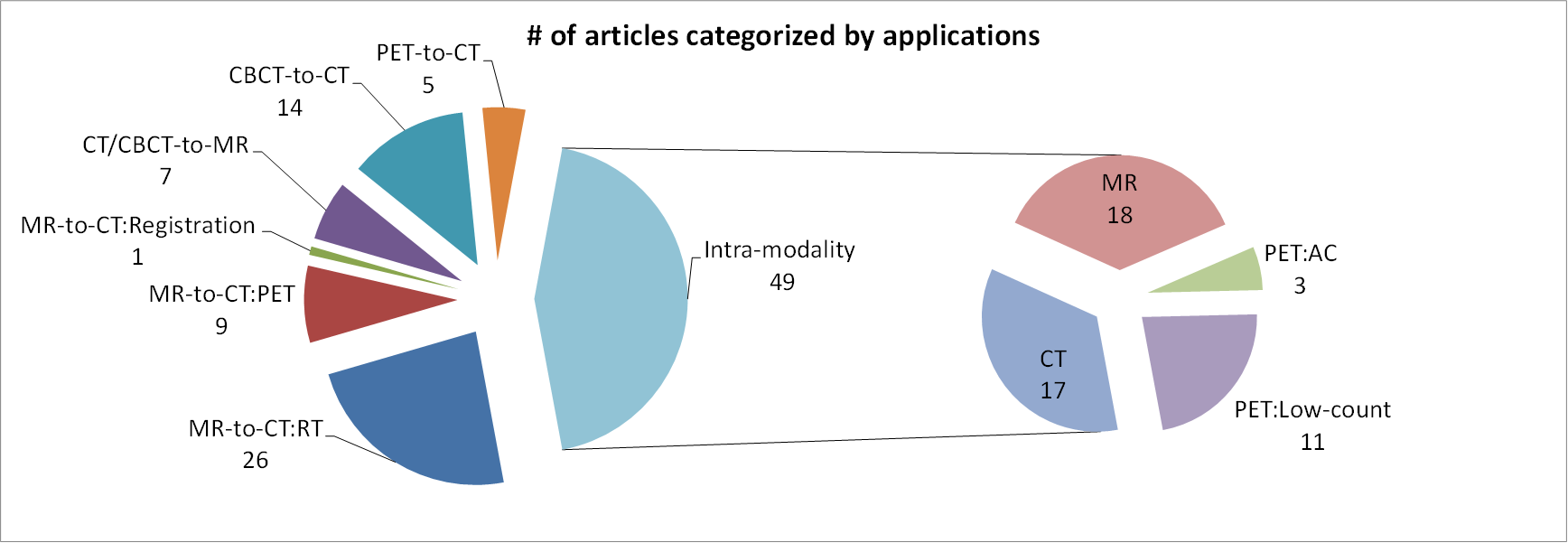}

\noindent Figure 3. Pie chart of numbers of articles in different categories of applications. MR-to-CT: RT, MR-to-CT: PET and MR-to-CT: Registration represent MR to CT image synthesis used in radiotherapy, PET and image registration, respectively. PET: AC and PET: Low-count represent PET image synthesis used in attenuation correction and low-count to full-count, respectively. 
\end{figure}

\bigbreak
\noindent 
\subsection{Inter-modality}

The group of inter-modality includes the studies of image synthesis from one image modality to a different one, such as from MR to CT, from CT to MR, from PET to CT, and etc. We also consider the transformation between CT and CBCT as inter-modality since they are acquired from different machines with different hardware, and are reconstructed with different principles and algorithms. Based on the studies image modalities, studies in this group were further divided into 4 subgroups, including MR-to-CT, CT/CBCT-to-MR, CBCT-to-CT and PET-to-CT. As shown in Figure 3, MR-to-CT synthesis, including its applications in radiation therapy, PET and image registration, accounts for about 2/3 of all studies and more than half in inter-modality studies. 

\bigbreak
\noindent 
{\bf 4.1.1 MR-to-CT}

Image synthesis from MR to CT is one of the first applications that utilize deep learning for medical image synthesis, and remains the most common topic in this field. Motivated by its success, a variety of applications aiming at transformation among other imaging modalities has been actively investigated. The main clinical motivation of MR-based synthetic CT is to replace CT by MR acquisition.\cite{RN1584} The image quality and appearance of the synthetic CT  in current studies are still considerably different from real CT, which prevents it from direct diagnostic usage. However, many studies demonstrated its utility for non- or indirect diagnostic purpose, such as treatment planning for radiation therapy and PET attenuation correction. 
 \bigbreak
       In current radiation therapy workflow, both MRI and CT are performed on patients for treatment planning. MR images feature excellent soft tissue contrast that is useful for delineation of the tumor and organs at risk (OARs),\cite{RN5498} while CT images provide electron density maps for dose calculation and reference images for pre-treatment positioning. The contours from MR images are propagated to CT images by image registration for treatment planning. However, using both imaging modalities not only leads to additional cost and time for the patient but also introduces systematic positioning errors up to 2 mm during the CT-MRI image fusion process.\cite{RN5499, RN5500, RN5501} Moreover, CT scan also introduces non-negligible ionization dose to patients,\cite{RN5502} especially those requiring re-simulation. Thus, it is highly desirable to bypass CT scans with a solely MRI-based treatment planning workflow. Emerging MR-Linac technology also motivates the exclusive use of MRI in radiotherapy.\cite{RN5504, RN5503} MR cannot replace CT in radiotherapy since the signal of MR images is from hydrogen nucleus, thus cannot provide material attenuation coefficients for electron density calibration and subsequent dose calculation.  
\bigbreak
       Replacing CT with MR is also preferable in current PET imaging although CT is widely combined with PET in order to perform both imaging exams serially on the same table. The CT images are used to derive the 511 keV linear attenuation coefficient map to model photon attenuation by a piecewise linear scaling algorithm \cite{RN1167, RN1759}. The linear attenuation coefficient map is then used to correct for the loss of annihilation photons by attenuation processes in the object on the PET images to achieve a satisfactory image quality. MR has been proposed to be incorporated with PET as a promising alterative to existing PET/CT system for its advantages of superior soft tissue contrast and radiation dose-free, with a similar challenge as in radiation therapy that MR images cannot be directly used to derive the 511 keV attenuation coefficients for attenuation correction process. Therefore, MR-to-CT image synthesis could be used in PET/MR system for photon attenuation correction. 
\bigbreak
       The missing of a one-to-one relationship between MR voxel intensity and CT HU values leads to a huge difference in image appearance and contrast, which makes intensity-based calibration methods fail. For example, bone is bright and air is dark on CT, while both are dark on MR. Conventional methods proposed in literatures either segment the MR images into several types of materials and then assign corresponding CT HU numbers,\cite{RN5510, RN5509, RN5506, RN5508, RN5507, RN5505} or register the MR images with an atlas with known CT HU numbers.\cite{RN5511, RN5512, RN5513} These methods heavily rely on the performance of segmentation and registration, which is always challenging due to the ambiguous air/bone boundary and large inter-patient variation, respectively. 
\bigbreak
       Table I and II listed the studies that synthesized CT images from MR for radiation therapy and PET attenuation correction, respectively. For synthetic CT in radiation therapy, the MAE is the most common and well-defined metrics by which almost every study reported the image quality of its synthetic CT. For synthetic CT in PET AC, the quality of PET with attenuation corrected by synthetic CT is more evaluated than the synthetic CT itself. For studies which presented several variants of methods, we listed the one with best MAE for radiation therapy, and best PET quality for PET AC. 

\bigbreak
\noindent 
{\bf 4.1.1.1 Synthetic CT quality}

In most of the studies, the MAE of the synthetic CT within patient body in about 40 to 70HU. Some of the reported results are comparable to the typical uncertainties observed in CT simulation. For example, the MAE of soft tissue reported in ref \cite{RN5145, RN4885, RN5456, RN5204, RN4879, RN4758, RN4787} are less than 40 HU. In contrast, the MAE of bone or air is more than 100 HU. The relatively poor performance on bone and air is expected due to their indistinguishable contrast on MR images. Another possible reason can be the misalignment between the CT and MR images in the patient datasets. The misalignment, which mostly happens on the bone, not only cause the intensity mapping error during the training process, but also leads to overestimation of error in the evaluation study since the error from misalignment was counted as synthesis error. Two studies also reported much higher MAE of rectum (~70HU) than soft tissue,\cite{RN5166, RN4787} which may also be attributed to its mismatch on CT and MR due to different filling status. Moreover, considering that the number of bone pixels are much less than those of soft tissue in patient body, the training process tends to map pixels to low HU region in the prediction stage. Potential solutions can be assigning higher loss weights on bone or adding bone-only images for training.\cite{RN4879} 
\bigbreak
       Compared with the conventional methods, learning-based methods demonstrated superior performance in synthetic CT accuracy in multiple studies, which indicates the advantage of the data-driven approaches over model-based methods. \cite{RN5166, RN5145, RN5023, RN5196} For example, the synthetic CT by atlas-based method was shown to be noisier and prone to error of registration, which led to significantly higher MAE than learning-based methods. However, atlas-based methods were shown to be more robust than learning-based methods to image quality variation in some cases.\cite{RN5166} One of the limitations of learning-based method is that its performance can be unpredictable when applied on datasets that are very different from the training datasets. The difference may come from abnormal anatomy, images with degraded quality due to severe artifacts and noise. Atlas-based methods, on the other hand, generate a weighted average of templates from its prior knowledge, thus are less likely to completely fail on unexpected cases.  
\bigbreak
       The reported results among these studies cannot be fairly compared because of different datasets, training and testing strategy, and etc. Thus, it is difficult to conclude the best method of performance. Some studies compared their proposed methods with other competing methods using same datasets, which may reveal the advantages and limitations of those methods. For example, a GAN-based method was shown to preserve better details, be more similar to real CT, and less noisy than a CNN-based method on a cohort of 15 brain cancer patients.\cite{RN5204} Specifically, GAN-based synthetic CT was more accurate at bone/air interfaces and fine structures, with around 10 HU less in MAE.  Largent \textit{et al.} compared U-net and GAN with different loss functions on 39 prostate patients: U-net with L2-norm loss, U-net with single-scale perceptual loss, GAN with L2 loss, GAN with single-scale perceptual loss, GAN with multiscale perceptual loss, and GAN with weighted multiscale perceptual loss. Quantitative results showed that the U-net methods had significantly higher MAE than their GAN counterparts. The perceptual loss in U-net and GAN did not help decrease MAE, nor provide any benefits for dose calculation accuracy. Lei \textit{et al.} compared CycleGAN and GAN-based method on brain and prostate cancer patients. Significant improvement of MAE was observed on CycleGAN results, with better visual results on fine structural details and contrast. CycleGAN results, which were less sensitive to local mismatch in the training CT/MR pairs, have less blurry bone boundaries than GAN results. Similar comparison results can also be found in the study of Liu \textit{et al.} where CycleGAN and GAN were compared on liver SBRT cases.\cite{RN4888} However, the dosimetry study showed minimal difference, which can be contributed to the VMAT plans that are not sensitive to HU inaccuracy.
 \bigbreak
       Among the reviewed studies, different types of MR sequences have been adopted for synthetic CT generation. The specific MR sequence used in each study usually depends on the accessibility. The optimal sequence yielding the best performance has not been studied. T1-weighted and T2-weighted sequences are two of the most common MR sequences used in diagnosis. Due to their wide availability, learning model can be trained from a relatively large number of datasets with CT and co-registered T1- or T2-weighted MR images. T2-w images may be preferable than T1-w since they intrinsically have better geometric accuracy in regions where subject-induced susceptibility is large, such as nasal cavity, and have less chemical shift artifacts at fat and tissue boundaries. However, air and bone have little contrast in either T1- or T2-weighted MR images, which may impede the extraction of the corresponding features in learning-based methods. Two-point Dixon sequence can separate water and fat, which is suitable for segmentation. It has already been applied in commercial PET/MR scanner with combination of volume-interpolated breath-hold examination (VIBE) for Dixon-based soft-tissue and air segmentation for PET AC as a clinical standard.\cite{RN1830, RN1831} Its drawback is again the poor contrast of bone, which results in the misclassification of bone as fat. In order to enhance the bone contrast to facilitate the feature extraction in learning-based methods, ultrashort echo time (UTE)– and/or zero echo time (ZTE) MR sequences have been recently used due to its capability to generate positive image contrast from bone.\cite{RN5002} Ladefoged \textit{et al.} and Blanc-Durand \textit{et al.} demonstrated the feasibility of UTE and ZTE MR sequences using U-net in PET/MR AC, repectively.\cite{RN1815, RN5088} However, neither of the two studies compared the using of UTE/ZTE and conventional MR sequence under the same deep learning network. Thus, the advantage of this specialized sequence has not been validated. Moreover, compared with conventional T1-/T2-weighted MR images, the UTE/ZTE MR images have little diagnostic value on soft tissue while have a long acquisition time, which may hinder its utility in time-sensitive cases such as whole-body PET/MR scans.  
\bigbreak
       Other studies attempted to use multiple MR images with different contrast as input in training and prediction since it is believed to be superior to single MR sequence in synthetic CT accuracy by providing additional features to the network. Qi \textit{et al.} proposed to use a 4-channel input that includes T1w, T2w, contrast-enhanced T1w, and contrast-enhanced T1w Dixon water images. Compared with the results from fewer channels, 4-channel result has lower MAE.\cite{RN4607} Florkow \textit{et al.} investigated single and multi-channel input using magnitude MR images and Dixon reconstructed water, fat, in-phase and opposed-phase images obtained from a single T1w multi-echo gradient-echo acquisition.\cite{RN4763} They found multi-channel input is able to improve synthetic CT generation than single-channel input. Among the multi-channel input configurations that they tested, the Dixon input outperformed the others. Tie \textit{et al.} used T2-w and pre- and post-contrast T1w MR images in a multi-channel multi-path architecture, and showed a significant improvement over multi-channel single-path and single-channel results.\cite{RN4612} An attractive combination is UTE/ZTE and Dixon, which provide contrast of bone against air and fat against soft tissue, respectively. \cite{RN1816, RN4993} Leynes \textit{et al.} showed that the synthetic CT using ZTE and Dixon MR has less error than that using Dixon alone.\cite{RN4993} Although the image quality improvement has been validated, the necessity of performing additional MR sequences for synthetic CT generation needs to be further evaluated in specific applications since it usually requires extra cost and acquisition time. 
\bigbreak
       In the reviewed studies, the CT and MR images in the training datasets were acquired separately on different machines. Thus, image registration is required between the CT and MR to create CT-MR pairs for training. The registration error is minimal at brain, and starts becoming an issue at pelvis due to different filling status in bladder and rectum, then be challenging in abdomen due to the variation introduced by respiratory and peristalsis. U-net and GAN-based methods are susceptible to registration error if using a pixel-to-pixel loss. Kazemifar \textit{et al.} proposed a possible solution that uses mutual information as the loss function in the generator of GAN to bypass the registration step in the training.\cite{RN4941} As CycleGAN was originally developed for unpaired image-to-image translation, CycleGAN-based methods feature higher robustness to registration error since it introduces cycle consistence loss to enforce the structural consistency between original one and cycle one, (e.g., force cycle MRI generated from synthetic CT to be the same as original MRI).\cite{RN4754, RN5369, RN4770, RN5364} 

\bigbreak
\noindent 
{\bf 4.1.1.2 MR-only radiation therapy}

For the studies aiming for radiation therapy, many of them evaluated the dosimetry accuracy of synthetic CT by calculating radiation treatment dose using same treatment plan and comparing it with that on real CT as ground truth. It is shown that the dose difference is about 1\%, which is small when compared with the current total uncertainty of dose delivery on patient (5\%) during the entire radiation therapy pathway. Compared to the large improvement of image accuracy, the improvement from learning-based methods over conventional methods in dosimetry accuracy on photon radiation therapy is relatively small. The conclusion about the significance of the improvement is also mixed.\cite{RN5166, RN5145} A potential reason is that the dose calculation on photon plans is quite forgiving to image inaccuracy, especially at homogeneous regions such as brain. For the widely studied volumetric modulated arc therapy (VMAT), the contribution from errors in images to dose also tends to cancel out in an arc. However, the small dosimetric improvement may still be worthwhile in cases such as SRS and SBRT where a large amount of dose is to be delivered into a small volume. In such cases, the dose calculation accuracy could be sensitive to the errors on synthetic CT around the target volume.\cite{RN1679} The recent adoption of non-coplanar beams may also be challenging to MR-based synthetic CT since the beam path length can be sensitive to the prediction error of patient surface due to the beam obliquity, which is worth further investigation. 

\bigbreak
       Studies have also evaluated the synthetic CT in the context of proton radiation treatment for prostate, liver and brain cancer.\cite{RN4793, RN4904, RN4617} Unlike photon, proton beams deposit dose with a very high dose gradient at the distal end of the beam. The proton treatment plan thus has highly conformal dose distribution to the target by proton beams coming from several angles. The local HU inaccuracy along the beam path on the planning CT would lead to shift of the highly conformal high-dose area, which may cause tumor to be substantially under-dosed or the organs-at-risk to be over-dosed.\cite{RN5281}  As shown in the Figure 4 in ref \cite{RN4904}, most dose difference of using synthetic CT was at the distal end of the proton beam. As reported by Liu \textit{et al.},\cite{RN4793, RN4904} the largest and mean absolute range difference is 0.56 cm and 0.19 cm among their 21 liver cancer patients, and 0.75 cm and 0.23 cm among 17 prostate cancer patients. 
\bigbreak
       In addition to dosimetry accuracy for treatment planning, another important aspect for the evaluation of synthetic CT is its geometry fidelity for patient setup. Unfortunately, the studies on synthetic CT positioning accuracy are sparse. Fu \textit{et al.} conducted patient alignment test by rigidly aligning the synthetic CT and real CT to the CBCT acquired at the first fraction.\cite{RN4879} The translation vector distance and absolute Euler angle difference between the two alignments were found to be less than 0.6mm and 0.5\si{\degree} on average, respectively. Gupta \textit{et al.} adopted a similar study, and found the translation difference was less than 0.7mm in one direction.\cite{RN4758} Apart from alignment with CBCT, the alignment between the derived DRR from synthetic CT and kV image of patient is also of clinical interest. However, no study on DRR alignment accuracy is found in the reviewed literatures. Note that the geometry accuracy of synthetic CT is not only affected by the performance of methods, but also the geometric distortion on MR images caused by magnetic field inhomogeneity as well as subject-induced susceptibility and chemical shift. Methods to mitigate the MR distortion are also important in improving synthetic CT accuracy in patient positioning. 

\bigbreak
\noindent 
{\bf 4.1.1.3 PET attenuation correction}

For the studies aiming for PET attenuation correction, the bias on PET quantification caused by the synthetic CT error has been evaluated. Although it is difficult to specify the tolerance level of quantification error before it affect clinician’s judgment, the general consensus is that quantification errors of 10\% or less typically do not affect diagnosis.\cite{RN1174} Based on the average relative bias represented by these studies, almost all of the proposed methods in the studies met this criterion. However, it should be noted that due to the variation among study objects, the bias in some volume-of-interests (VOIs) of some patients may exceed 10\%.\cite{RN1815, RN4993} It suggests that special attention should be given to the standard deviation of bias as well as its mean when interpreting results since the proposed methods may have poor local performance that would affect some patients. On the other hand, listing or plotting the results of every data points, or at least the range, instead of simply giving a mean$\mathrm{\pm}$STD in presenting results, would be more informative in demonstrating the performance of the proposed methods. 
\bigbreak
       Since bone has the highest attenuation capability due to its high density and atomic number,\cite{RN1864} its accuracy on synthetic CT plays a vital role in the final accuracy of the attenuation corrected PET. Compared with the evaluation for radiation therapy, the bias and geometry accuracy of bone on the synthetic CT is more often evaluated for PET AC. Multiple studies showed that synthetic CT with better bone accuracy tends to generate more accurate PET globally.\cite{RN4866, RN1815, RN1816, RN5172} The more accurate synthetic CT images by learning-based methods than conventional methods also lead to a more accurate PET AC. Such improvements were found to be significant in the reviewed studies. It was shown that PET AC by conventional synthetic CT methods have about 5\% bias on average among selected VOIs, while for learning-based methods, the bias was reduced to around 2\%.\cite{RN1816, RN5224, RN4993, RN5002, RN5172}  
\bigbreak
       In addition to the two widely studied applications, i.e. radiation treatment simulation and PET AC, using synthetic CT from MR to aid intra-modality image registration has been proved promising. Direct image registration between CT and MR is very challenging due to the distinct image contrast, and can be unreliable in deformable registration algorithms where large distortion is allowed. McKenzie \textit{et al.} proposed a CycleGAN-based network to generate synthetic CT, and used the synthetic CT to replace the MR in MR-CT registration for head-and-neck.\cite{RN4681} In this way, the multimodality registration problem is reduced to a mono-modality one. In their study as summarized in Table III, it was found that with the same deformable registration algorithm, the average landmark error decreased from 9.8$\mathrm{\pm}$3.1 mm in direct MR-CT registration to 6.0$\mathrm{\pm}$2.1 mm in using synthetic CT as bridge. Similar results were also found in the registration at CT-MR direction.  

\bigbreak
\noindent \textbf{Table I.} Summary of studies on MR-based synthetic CT for radiation therapy

\begin{longtable}{|p{0.5in}|p{0.8in}|p{1.1in}|p{0.8in}|p{1.0in}|p{0.8in}|} \hline 
\textbf{Network} & \textbf{MR parameters} & \textbf{Site, and \# of patients in training/testing} & \textbf{Key findings in image quality} & \textbf{Key findings in dosimetry} & \textbf{Author, year} \\ \hline 
U-net & 1.5T T1w without contrast & Brain: 18, 6-fold cross validation & MAE (HU): 84.8$\mathrm{\pm}$17.3  & N/A${}^{*}$ & Han, 2017 \cite{RN5023} \\ \hline 
GAN & N/A & Brain: 16\newline Pelvis: 22 & MAE (HU): 92.5$\mathrm{\pm}$13.9 & N/A & Nie \textit{et al.}, 2018\cite{RN5196} \\ \hline 
CNN & T1w & Brain: 16, leave-one-out\newline Pelvis: 22, leave-one-out & MAE (HU): 85.4$\mathrm{\pm}$9.24 (brain)\newline 42.4$\mathrm{\pm}$5.1 (Pelvis) & N/A & Xiang \textit{et al.}, 2018\cite{RN5493} \\ \hline 
CNN & 1.5T T1w & Brain: 52, 2-fold cross validation & MAE (HU): 67$\mathrm{\pm}$11 & Dose difference $\mathrm{<}$1\% & Dinkla \textit{et al.}, 2018\cite{RN5456} \\ \hline 
U-net & 3T T2w & Pelvis: 39, 4-fold cross validation & MAE (HU): 32.7$\mathrm{\pm}$7.9 & Dose difference $\mathrm{<}$1\% & Arabi \textit{et al.}, 2018 \cite{RN5166} \\ \hline 
U-net & 3T T2w & Pelvis: 36 training/15 testing & MAE (HU): 29.96$\mathrm{\pm}$4.87 & Dose difference of max dose in PTV $\mathrm{<}$1.01\% & Chen \textit{et al.}, 2018 \cite{RN5145} \\ \hline 
GAN & 1T post-Gadolinium T1w  & Brain: 15, 5-fold cross validation & MAE (HU): 89.3$\mathrm{\pm}$10.3 & N/A & Emami \textit{et al.}, 2018\cite{RN5204} \\ \hline 
GAN & Dixon in-phase, fat and water & Pelvis: 91 (59 prostate+18 rectal+14 cervical cancer), 32 (prostate) training/59 (rest) testing & MAE (HU): 65$\mathrm{\pm}$10 (Prostate)\newline 56$\mathrm{\pm}$5 (Rectum)\newline 59$\mathrm{\pm}$6 (Cervix) & Dose difference $\mathrm{<}$ 1.6\% & Maspero \textit{et al.}, 2018 \cite{RN5180} \\ \hline 
U-net & 3T in-phase Dixon T2w & Head and neck: 22 training/12 testing & MAE (HU): 75$\mathrm{\pm}$9 & Mead dose difference -0.03\%$\mathrm{\pm}$0.05\% overall,\newline -0.07\%$\mathrm{\pm}$0.22\% in $\mathrm{>}$90\% of prescription dose volume & Dinkla \textit{et al.}, 2019\cite{RN4885} \\ \hline 
U-net & 1.5T T1w without contrast & Pelvis: 20, 5-fold cross validation & MAE (HU): 40.5$\mathrm{\pm}$5.4 (2D)\newline 37.6$\mathrm{\pm}$5.1 (3D) & N/A & Fu \textit{et al.}, 2019\cite{RN4879} \\ \hline 
U-net & 3T in-phase Dixon T1w & Brain: 47 training/13 testing & MAE (HU): 17.6$\mathrm{\pm}$3.4 & Mean target dose difference 2.3$\mathrm{\pm}$0.1\% & Gupta \textit{et al.}, 2019\cite{RN4758} \\ \hline 
GAN & 1.5T post-Gadolinium T1w & Brain: 77, 70\% training/12\% validation/18\% testing & MAE (HU):\newline 47.2$\mathrm{\pm}$11.0 & Mean DVH metrics difference $\mathrm{<}$1\% & Kazemifar \textit{et al.}, 2019\cite{RN4941} \\ \hline 
GAN & 3T T2w & Pelvis: 39, training/testing: 25/14, 25/14, 25/11 & MAE (HU): 34.1$\mathrm{\pm}$7.5 & PTV V95\% difference $\mathrm{<}$ 0.6\% & Largent \textit{et al.}, 2019\cite{RN4787} \\ \hline 
CycleGAN & Brain: T1w\newline Pelvis: T2w & Brain: 24\newline Pelvis: 20\newline Leave-one-out cross validation & MAE (HU): 55.7$\mathrm{\pm}$9.4 (Brain)\newline 50.8$\mathrm{\pm}$15.5 (Pelvis) & N/A & Lei \textit{et al.}, 2019\cite{RN4913} \\ \hline 
U-net & 1.5T T1w & Brain: 30 training/10 testing & MAE (HU): 75$\mathrm{\pm}$23 & PTV V95\% difference 0.27\%$\mathrm{\pm}$0.79\% & Liu \textit{et al.}, 2019\cite{RN4973} \\ \hline 
CycleGAN & 3T/1.5T T1w & Liver: 21, leave-one-out cross validation & MAE (HU): 72.87$\mathrm{\pm}$18.16 & Mean DVH metrics difference $\mathrm{<}$1\% for both photon and proton plans & Liu \textit{et al.}, 2019\cite{RN4888} and Liu \textit{et al.}, 2019\cite{RN4904} \\ \hline 
CycleGAN & 1.5T T2w & Pelvis: 17, leave-one-out cross validation & MAE (HU): 51.32$\mathrm{\pm}$16.91 & Mean DVH metrics difference $\mathrm{<}$1\% (Proton plan) & Liu \textit{et al.}, 2019\cite{RN4904} \\ \hline 
U-net & 1.5T T1w & Brain: 57 training/28 validation/4 testing & MAE (HU): (82, 147)${}^{+}$ & Gamma passing rate: $\mathrm{>}$95\% at (1\%, 1mm) for photon plan,\newline $\mathrm{>}$90\% at (2\%, 2mm) for proton plan & Neppl \textit{et al.}, 2019\cite{RN4865} \\ \hline 
GAN & 0.35T T1w & Breast: 48 training/12 testing & MAE (HU): 16.1$\mathrm{\pm}$3.5 & PTV D95 difference$\mathrm{<}$1\% & Olberg \textit{et al.}, 2019\cite{RN4842} \\ \hline 
CycleGAN & 1.5T T1w & Brain: 50 & MAE (HU): 54.55$\mathrm{\pm}$6.81 & PTV D95 difference$\mathrm{<}$0.5\% (proton plan) & Shafai-Erfani \textit{et al.}, 2019\cite{RN4617} \\ \hline 
U-net & 1.5T T2w & Head and neck: 23 training/10 testing & MAE (HU): 131$\mathrm{\pm}$24 & N/A & Wang \textit{et al.}, 2019\cite{RN4685} \\ \hline 
U-net & 3T T1w Dixon & Pelvis: 27, 3-fold cross validation & MAE (HU): (33, 40) & N/A & Florkow \textit{et al.}, 2020\cite{RN4763} \\ \hline 
GAN & T1w+T2w+ FLAIR & Brain: 15 & MAE (HU): 108.1$\mathrm{\pm}$24.0 & DVH metrics difference $\mathrm{<}$ 1\% & Koike \textit{et al.}, 2020\cite{RN4699} \\ \hline 
GAN & T1w+T2w+ Contrast-enhanced T1w+ Contrast-enhanced T1w Dixon water & Head and neck: 30 training/15 testing & MAE (HU): 69.98$\mathrm{\pm}$12.02 & Mean average dose difference $\mathrm{<}$1\% & Qi \textit{et al.}, 2020\cite{RN4607} \\ \hline 
GAN & 1.5T Pre contrast T1w+post contrast T1w+T2w & Head and neck: 32, 8-fold cross validation & MAE (HU): 75.7$\mathrm{\pm}$14.6 & N/A & Tie \textit{et al.}, 2020\cite{RN4612} \\ \hline 
GAN & 1.5T and 3T T2w from three scanners & Pelvis: 11 training from two scanner/8 testing from one scanner & MAE (HU): 48.5$\mathrm{\pm}$6 & Maximum dose difference in target = 1.3\% & Boni \textit{et al.}, 2020 \cite{RN5497} \\ \hline 
\end{longtable}

*N/A: not available, i.e. not explicitly indicated in the publication

${}^{+}$Numbers in parentheses indicate minimum and maximum values.

\bigbreak
\textbf{Table II.} Summary of studies on MR-based synthetic CT for PET attenuation correction.

\begin{longtable}{|p{0.7in}|p{0.8in}|p{0.9in}|p{0.9in}|p{0.8in}|p{0.9in}|} \hline 
\textbf{Network} & \textbf{MR parameters} & \textbf{Site, and \# of patients in training/testing} & \textbf{Key findings in synthetic CT quality} & \textbf{Key findings in PET quality} & \textbf{Author, year} \\ \hline 
U-net & Dixon and ZTE & Brain: 14, leave-two-out & MAE (\%): 12.62$\mathrm{\pm}$1.46 & Absolute bias $\mathrm{<}$3\% among 8 VOIs & Gong \textit{et al.}, 2018\cite{RN1816} \\ \hline 
U-net (Encoder-decoder) & 3T UTE & Brain: 30 pre-training/6 training/8 testing & N/A* & Bias (\%): -0.8$\mathrm{\pm}$0.8 to 1.1$\mathrm{\pm}$1.3 among 23 VOIs & Jiang \textit{et al.}, 2018\cite{RN5224} \\ \hline 
U-net & 3T Dixon and ZTE & Pelvis:26, 10 training/16 testing & Mean error (HU): -12$\mathrm{\pm}$78 & RMSE (\%): 2.68 among 30 bone lesions, 4.07 among 60 soft-tissue lesions & Leynes \textit{et al.}, 2018\cite{RN4993} \\ \hline 
U-net (Encoder-decoder) & 1.5T T1w & Brain: 30 training/10 testing & N/A & Bias (\%): -3.2$\mathrm{\pm}$1.3 to 0.4$\mathrm{\pm}$0.8\textbf{} & Liu \textit{et al.}, 2018\cite{RN5002} \\ \hline 
U-net & 1.5T T1w & Brain: 44 training/11 validation/11 testing & Global Bias (\%): -1.06$\mathrm{\pm}$0.81 & Global Bias(\%): -0.49$\mathrm{\pm}$1.7 for 11C-WAY-100635\newline -1.52$\mathrm{\pm}$0.73 for 11C-DASB & Spuhler \textit{et al.}, 2019\cite{RN5173} \\ \hline 
U-net & Dixon-VIBE & Pelvis: 28 pairs from 19 patients, 4-fold cross validation & MAE (\%): 2.36$\mathrm{\pm}$3.15 & Bias (\%): 0.27$\mathrm{\pm}$2.59 for fat\newline -0.03$\mathrm{\pm}$2.98 for soft tissue\newline -0.95$\mathrm{\pm}$5.09 for bone & Torrado-Carvajal \textit{et al.}, 2019\cite{RN5172} \\ \hline 
U-net & ZTE & Brain: 23 training/47 testing & N/A & Bias (\%): -1.8$\mathrm{\pm}$1.9 to 1.7$\mathrm{\pm}$2.6 among 70 VOIs & Blanc-Durand \textit{et al.}, 2019\cite{RN1815} \\ \hline 
U-net & UTE & Brain: 79 (pediatric), 4-fold cross validation & N/A & Bias (\%): -0.2 to 0.5 in 95\% CI & Ladefoged \textit{et al.}, 2019\cite{RN5088} \\ \hline 
GAN & 3T T1w & Brain: 40, 2-fold cross validation & MAE (HU): 302$\mathrm{\pm}$79 (bone) & Absolute bias $\mathrm{<}$ 4\% among 63 VOIs & Arabi \textit{et al.}, 2019\cite{RN4866} \\ \hline 
\end{longtable}

*N/A: not available, i.e. not explicitly indicated in the publication
\bigbreak

\noindent \textbf{Table III.} Summary of studies on MR-based synthetic CT for registration.

\begin{longtable}{|p{0.7in}|p{0.8in}|p{0.9in}|p{0.9in}|p{0.8in}|p{0.9in}|} \hline 
\textbf{Network} & \textbf{MR parameters} & \textbf{Site, and \# of patients in training/testing} & \textbf{Key findings in synthetic CT quality} & \textbf{Key findings in registration accuracy} & \textbf{Author, year} \\ \hline 
CycleGAN & 0.35T  & Head and neck: 25, 5-fold cross validation & N/A* & landmark error (mm): 6.0$\mathrm{\pm}$2.1 (MR-to-CT)\newline 6.6$\mathrm{\pm}$2.0 (CT-to-MR) & McKenzie \textit{et al.}, 2019\cite{RN4681} \\ \hline 
\end{longtable}

*N/A: not available, i.e. not explicitly indicated in the publication

\bigbreak
\noindent 
{\bf 4.1.2 CT/CBCT-to-MRI}

Due to the superior soft tissue contrast on MRI, it is attractive to generate synthetic MRI from CT or CBCT in applications that are sensitive to soft tissue contrast, such as segmentation.\cite{RN1442} Comparing with synthesizing CT from MR, synthesizing MR from CT/CBCT seems more challenging since MR contains much more contrast and details that need to be recovered but are not shown on CT images. Deep learning methods, however, are quite competent in mapping high non-linearity, which makes the proposed application possible. 
\bigbreak
       As listed in Table IV, the related studies in the reviewed literatures adopted similar networks as MR-to-CT synthesis. In most studies, the generated synthetic MR served as a bridge that is used in other applications, thus the image intensity accuracy of synthetic MR was not reported. In studies that reported synthetic MR accuracy, MAE is less meaningful than other image similarity metrics such as PSNR since the MR image intensity is relative. 
\bigbreak
       Jiang \textit{et al.} proposed to use synthetic MR to augment the training data for MR tumor segmentation in lung.\cite{RN4862} In their study, 81 MR image sets have tumor contours delineated by experts, which was considered as a small data size for training a segmentation model. In order to enlarge the training datasets, they employed a GAN-based model to generate synthetic MR from 377 CT image sets which has tumor labelled using other groups of unpaired MR image sets. The 377 synthetic MR image sets with tumor labels were then incorporated into segmentation model training. The addition of synthetic MR in training dataset was shown to be beneficial in improving the tumor segmentation performance. It increased the DSC of tumor to 0.75$\mathrm{\pm}$0.12 from 0.50$\mathrm{\pm}$0.26 in which synthetic MR was not included in training datasets. The study also showed that among the synthetic MRIs generated by different methods, the ones closer to real MR enabled better segmentation results. 
\bigbreak
        In the training stage of segmentation model in the above study, the training target contours for the synthetic MRIs were not delineated based on MR but on CT. Thus, in the testing stage, the output contours were also expected to be CT-based rather than MR-based. Since the delineation of tumor relies on the image contrast, the contour for a same object is usually different on CT and MRI. In some cases, contours from MR is more accepted as golden standard due to its superior soft tissue contrast than those on CT. Using CT-based contours as training target for synthetic MR may not only confuse the network, but also waste the superior soft tissue contrast of MRIs.  
\bigbreak
       In the studies of Dong \textit{et al.} and Lei \textit{et al.}, the synthetic MRIs were used as a bridge to facilitate segmentation on CT/CBCT images.\cite{RN4750, RN4684, RN5301} The segmentation targets in their study include prostate, which has low contrast on CT/CBCT but high on MRI images, and tends to be over-contoured with larger variation on CT/CBCT images when compared with using MRI only or CT+MRI.\cite{RN5461, RN5460} The synthetic MRIs generated by CT were then aimed at providing superior soft tissue contrast for prostate segmentation. In their studies, paired CT and MRI image sets were used, and the prostate contours used as training targets and ground truth for synthetic MR were delineated on MR or both CT and MRI. It shows that the mean DSC of prostate between segmentation results and ground truth increased from 0.82$\mathrm{\pm}$0.09 of direct segmentation on CT to 0.87$\mathrm{\pm}$0.04 of segmentation on synthetic MR with statistical significance. 
\bigbreak
\noindent \textbf{Table IV.} Summary of studies on CT/CBCT-based synthetic MR.

\begin{longtable}{|p{0.7in}|p{0.8in}|p{0.9in}|p{0.9in}|p{0.8in}|p{0.9in}|} \hline 
\textbf{Network} & \textbf{MR parameters} & \textbf{Site, and \# of patients in training/testing} & \textbf{Key findings in synthetic MR quality} & \textbf{Application} & \textbf{Author, year} \\ \hline 
CycleGAN & T2w & Pelvis: 140, 5-fold cross validation & N/A* & Male pelvis multi-organ segmentation on CT & Dong \textit{et al.}, 2019\cite{RN4750} \\ \hline 
GAN & 3T T2w & Lung: 42 MRIs and 377 CTs, unpaired training & Kullback--Leibler divergence in tumor: 0.069 & Augment training data for lung tumor segmentation on MR & Jiang \textit{et al.}, 2019\cite{RN4862} \\ \hline 
CycleGAN & 3T T2w & Spine: 549 training/92 testing & PSNR(dB): 64.553 $\mathrm{\pm}$1.890 & Diagnosis & Jin \textit{et al.}, 2019\cite{RN5457} \\ \hline 
CycleGAN & 3T T2w & Brain: 192 training/10 testing & PSNR(dB): 65.35 & Diagnosis & Jin \textit{et al.}, 2019\cite{RN5455} \\ \hline 
GAN & 1.5T and 3T T2w & Spine: 280 pairs in training/15 testing & PSNR(dB): 64.9$\mathrm{\pm}$1.86 & Diagnosis & Lee \textit{et al.}, 2020\cite{RN5458} \\ \hline 
CycleGAN & 1.5T T2w & Pelvis: 49, leave-one-out & N/A & Prostate segmentation on CT & Lei \textit{et al.}, 2020\cite{RN5301} \\ \hline 
CycleGAN & T2w & Pelvis: 100, 5-fold cross validation & N/A & Male pelvis multi-organ segmentation on CBCT & Lei \textit{et al.}, 2020\cite{RN4684} \\ \hline 
\end{longtable}

*N/A: not available, i.e. not explicitly indicated in the publication

\bigbreak
\noindent 
{\bf 4.1.3 CBCT-to-CT}

CBCT and CT share a common basic physics principle of X-ray attenuation and image reconstruction concept of back projection. However, they are different in the detailed implementation of acquisition and reconstruction, and clinical utility, thus they are considered as two image modalities in this review study. 
\bigbreak
       CBCT has been increasingly utilized in image-guided radiation therapy to improve treatment performance. CBCTs are acquired at the time of treatment delivery and provide detailed anatomic information in the treatment position. In clinical practice, CBCT is primarily used to determine the degree of patient setup error and inter-fraction motion by comparing the displacement of anatomic landmarks from the treatment planning CT images.\cite{RN1146} More demanding applications of CBCT have been proposed with the increasing use of adaptive radiation therapy, such as daily estimation of delivered dose based on CBCT images, and automatic contouring on CBCTs based on a deformable image registration (DIR) with the pCT.\cite{RN772, RN727} 
\bigbreak
       Unlike CT scanners using fan-shaped X-ray beam with multi-slice detectors, CBCT uses cone-shaped X-ray beam with a flat panel detector. The flat panel detector features a high spatial resolution and a wide coverage along the z-axis, but also gets more scatter signals since the scatter X-ray generated from the entire volume can reach the detector. The scatter signals cause severe streaking and cupping artifacts on the CBCT images and lead to significant CT number errors. Such errors complicate the calibration process of CBCT Hounsfield Unit (HU) to electron density for dose calculation.\cite{RN741} The degraded image contrast and the suppression of bone CT number can also cause large errors in DIR for contour propagation from planning CT to CBCT.\cite{RN773} The significantly degraded image quality of CBCT prevents it from those advanced quantitative usage in radiation therapy. 
\bigbreak
       Many correction methods for CBCT shading artifacts have been proposed, including hardware-based pre-processing methods\cite{RN731, RN732, RN733} and software-based post-processing techniques \cite{RN737, RN738, RN736, RN735, RN5514, RN734} These methods enhance the scatter correction performance, while their implementations entail combined considerations of computational complexity, imaging dose, scan time, practicality, and efficacy. With these correction methods, residual artifacts are still commonly observed in clinical CBCT images. Moreover, most of the existing methods cannot restore the true Hounsfield Unit (HU) value in CBCT images; i.e., the pixel values in CBCT images are not calibrated identically to planning CT images in the treatment planning system for dose calculation. 
\bigbreak
       Deep learning-based methods, as listed in Table V, have been proposed to correct and restore the CBCT HU values to be close to those of CT by exploiting its outstanding image translation abilities. CBCT images are reconstructed from hundreds of 2D projections from different angles. A few studies applied the neural network in the projection domain in order to enhance the quality of the projection images prior to volume reconstruction. More studies directly converted the reconstructed CBCT image volumes into high quality as CT. Projection-domain methods can be advantageous in a larger number of training projections ($>$300) than training images of image domain ($<$100) for each scan. Moreover, the appearances of cupping and streaking artifacts caused by scatter on CBCT images are more random and diverse among different patients than the distribution of scatter photons on projection images. In other words, the scatter presented on projection is more predictable and easier to learn for neural networks. The image-domain methods do not train models on non-anthropomorphic phantom since it is useless for patient image sets due to the huge difference in image features. However, such difference in image features is much less in projection domain. Nomura \textit{et al.} showed that the features characterizing scatter distribution in anthropomorphic phantom projections can be learned from non-anthropomorphic phantom projections.\cite{RN5470} The potential reason is that the neural network successfully learned the inherent relationship between the scatter distribution and objective thickness on projection domain. The relationship between the scatter artifacts and the objective appearance is much more complicated and thus can be hardly learned on image domain. 
\bigbreak
       In the reviewed studies, the learning target of CBCT images/projections is mostly the corresponding CT images/projections that captured on the same patient. However, mismatch is commonly seen between CT and CBCT, thus registration is required to reduce artifact caused by mismatch. Liu \textit{et al.} compared the performance of their method between using rigidly and deformably registered CBCT-CT training data in their pancreas study.\cite{RN5305} They found that synthetic CT by rigidly registered training data had slightly higher in MAE than deformably registered training data (58.45$\mathrm{\pm}$13.88 HU vs 56.89$\mathrm{\pm}$13.84 HU, p$>$0.05), and less noise with better organ boundaries. Kurz \textit{et al.} showed that using unmatched CT and CBCT as training datasets in CycleGAN without pixel-by-pixel loss function is feasible to generate synthetic CT with satisfactory quality.\cite{RN5467} To bypass the registration step, Hansen \textit{et al.} and Landry \textit{et al.} proposed to correct CBCTs by conventional method first, and then use the corrected CBCTs as the learning target. Since the corrected CBCTs are always in the same geometry as original CBCT, registration is no longer needed.\cite{RN5469, RN5463} However, the quality of the synthetic CT is limited by the conventional method-generated CBCT. 
\bigbreak
       In studies that compared the performance of the proposed deep learning-based methods with conventional CBCT correction methods using same datasets, learning-based methods feature better image quality.\cite{RN5468, RN4884, RN5462, RN5470, RN5466} Adrian \textit{et al.} found their U-net-based method outperformed two conventional methods, deformable registration method and analytical image-based correction method, with the lowest MAE of synthetic CT, the lowest spatial non-uniformity, and the most accurate bone geometry.\cite{RN5468} Harms \textit{et al.} observed a lower noise of their synthetic CT and a more similar appearance as real CT when compared with a conventional image-based correction method.\cite{RN4884} Conventional correction methods are designed to enhance a specific aspect of image quality, while the learning-based methods, aiming to generate synthetic CT from CBCT, would change every aspect of image quality to be close to CT, such as noise level that is usually not considered in conventional correction methods. A few studies also compared different networks with same patient datasets, and it is shown that CycleGAN outperforms GAN and U-net. \cite{RN4915, RN5305} 
\bigbreak
       Synthetic CTs are found to have significant improvement over original CBCTs in dosimetry accuracy, and are close to planning CT for photon dose calculation. The feasibility in VMAT photon plans have been evaluated in various sites by investigating selected DVH metrics and dose difference/Gamma difference. Fig. 4 in the study of Liu \textit{et al.} demonstrates that large local dose calculation error happened at locations with severe artifacts, and synthetic CT successfully mitigated the artifacts and therefore the dosimetry error.\cite{RN5305}  Compared with photon, the dosimetry accuracy for proton plan is more challenging. Proton range shift on synthetic CT is usually about 3mm and can be up to 5mm. \cite{RN5469, RN5467, RN5463} 

\bigbreak
\noindent \textbf{Table V.} Summary of studies on CBCT-based synthetic CT for radiation therapy.

\begin{longtable}{|p{0.7in}|p{0.8in}|p{0.9in}|p{0.9in}|p{0.8in}|p{0.9in}|} \hline 
\textbf{Network} & \textbf{Projection or image domain} & \textbf{Site, and \# of patients in training/testing} & \textbf{Key findings in synthetic CBCT quality} & \textbf{Key findings in dosimetry} & \textbf{Author, year} \\ \hline 
U-net & Projection & Pelvis: 15 training/7 testing/8 evaluation & MAE (HU): 46 & Passing rate for 2\% dose difference: 100\% for photon plan,\newline 15\%-81\% for proton plan  & Hansen \textit{et al.}, 2018\cite{RN5469} \\ \hline 
U-net & Image & Pelvis: 20, 5-fold cross validation & PSNR (dB): 50.9 & N/A* & Kida \textit{et al.}, 2018\cite{RN5462} \\ \hline 
CNN & Image & Lung: 15 training/5 testing & PSNR (dB):8.823 & N/A & Xie \textit{et al.}, 2018\cite{RN5466} \\ \hline 
U-net & Image & Head and neck: 30 training/7 validation/7 testing\newline Pelvis: 6 training/5 testing & MAE (HU): 18.98 (head and neck)\newline 42.40 (pelvis) & N/A & Chen \textit{et al.}, 2019\cite{RN4682} \\ \hline 
CycleGAN & Image & Brain: 24, leave-one-out\newline Pelvis: 20, leave-one-out & MAE (HU): 13.0$\mathrm{\pm}$2.2 (brain)\newline 16.1$\mathrm{\pm}$4.5 (Pelvis) & N/A & Harms \textit{et al.}, 2019\cite{RN4884} \\ \hline 
CycleGAN & Image & Pelvis: 18 training/7 validation/8 testing & MAE (HU): 87 (79, 106)${}^{+}$ & Passing rate for 2\% dose difference: 100\% for photon plan,\newline 71\%-86\% for proton plan & Kurz \textit{et al.}, 2019\cite{RN5467} \\ \hline 
U-net & Image & Pelvis: 27 training/7 validation/8 testing & MAE (HU): 58 (49, 69) & Passing rate for 2\% dose difference: $\mathrm{>}$99.5\% for photon plan,\newline $\mathrm{>}$80\% for proton plan & Landry \textit{et al.}, 2019\cite{RN5463} \\ \hline 
U-net & Image & Head and neck: 50 training/10 validation/10 testing & MAE (HU): (6, 27)  & Average DVH metrics difference: 0.2$\mathrm{\pm}$0.6\% & Li \textit{et al.}, 2019\cite{RN5464} \\ \hline 
CycleGAN & Image & Head and neck: 81 training/9 validation/20 testing & MAE (HU): 29.85$\mathrm{\pm}$4.94 & Gamma passing rate at (1\%, 1mm): 96.26$\mathrm{\pm}$3.59\% & Liang \textit{et al.}, 2019\cite{RN4915} \\ \hline 
U-net & Projection & 1800 projections in training (simulation)/200 validation (simulation)/360 testing (phantom) & MAE (HU): 17.9$\mathrm{\pm}$5.7 & N/A & Nomura \textit{et al.}, 2019\cite{RN5470} \\ \hline 
U-net & Image & Head and neck: 33, 3-fold cross validation & MAE (HU): 36.3$\mathrm{\pm}$6.2 & Gamma passing rate at (2\%, 2mm): 93.75-99.75\% (proton) & Adrian \textit{et al.}, 2020\cite{RN5468} \\ \hline 
U-net & Image & Head and neck: 40 training/15 testing & MAE (HU): 49.28 & N/A & Yuan \textit{et al.}, 2020\cite{RN5465} \\ \hline 
CycleGAN & Image & Pelvis: 16 training/4 testing & Mean error (HU): (2, 14) & N/A & Kida \textit{et al.}, 2020\cite{RN5496} \\ \hline 
CycleGAN & Image & Pancreas: 30. Leave-one-out & MAE (HU): 56.89$\mathrm{\pm}$13.84 & DVH metrics difference $\mathrm{<}$ 1Gy & Liu \textit{et al.}, 2020\cite{RN5305} \\ \hline 
\end{longtable}

*N/A: not available, i.e. not explicitly indicated in the publication

${}^{+}$Numbers in parentheses indicate minimum and maximum values.

\bigbreak
\noindent 
{\bf 4.1.4 PET-to-CT}

In a PET-only scanner where neither CT nor MR is available, transmission scan with an external positron source rotated around the patient is currently used to determine the attenuation of patient body for attenuation correction. It is thus desirable to use the non-attenuation-corrected (NAC) PET to generate synthetic CT to provide anatomical information by the powerful image style transfer ability of deep learning methods. Moreover, for PET/MR, although MR provides anatomical images, the current atlas or registration-based methods in MR-based PET/MR attenuation correction are subject to errors in the bone on the derived attenuation map. Deriving attenuation map from existing NAC PET is therefore an attractive alternative.
\bigbreak
       The related studies are listed in Table VI. Similar to other synthetic image generation, the synthetic CT images were generated from the NAC PET images using the deep learning model trained by pairs of NAC PET and CT images that were acquired from a PET/CT scanner. Synthesizing CT from NAC PET images is intrinsically challenging since the NAC PET images have much lower spatial resolution than CT and provide little anatomical information. In the studies of Hwang \textit{et al.}, time-of-flight information was used to generate maximum-likelihood reconstruction of activity and attenuation maps as input since they provides more anatomical information than NAC PET.\cite{RN5256, RN5471} Despite of these difficulties, the reported average errors are all within the 10\% consensus tolerance, which is competitive to the results by MR-based synthetic CT. 
\bigbreak
       Whole-body PET scan has been an important imaging modality in finding tumor metastasis. Most of the reviewed studies about image synthesis in PET developed their proposed methods for brain applications. Although the learning-based methods are data-driven and not site-specific, they may not be readily applicable to the whole body due to the high anatomical heterogeneities, large variance of activity among different organs, and inter-subject variability. Hwang \textit{et al.} and Dong \textit{et al.} investigated learning-based whole-body PET AC using synthetic CT generation strategy.\cite{RN4754, RN5471} Both studies reported average bias on lesion to be around 1\%, which is promising for clinical use. Dong \textit{et al.} reported average bias within 5\% in all selected organs except $>$10\% in lungs in both studies. The authors attributed the poor performance on lung to the tissue inhomogeneity and insufficient representative training datasets. They also found that the synthetic CTs showed blurriness in lung like respiratory motion artifacts that were not shown on CT, which indicates that synthetic CTs are more matched with PETs than CT and can be more suitable for AC. Both studies are performed for PET-only scanner, and so far, there is no learning-based methods developed for PET/MR whole body scanner. Compared with PET-only scheme, the PET/MR provides the anatomical structural information from MR, while the integration of the additional MR into PET AC can be more challenging than brain scans since the MR may have a limited field of view (FOV), longer scan time that introduces more motion, and degraded image quality due to larger inhomogeneous-field region. 

\bigbreak
\noindent \textbf{Table VI.} Summary of studies on PET-based synthetic CT for PET attenuation correction.

\begin{longtable}{|p{0.5in}|p{1.2in}|p{1.2in}|p{1.2in}|p{0.9in}|} \hline 
\textbf{Network} & \textbf{Site, and \# of patients in training/testing} & \textbf{Key findings in synthetic CT quality} & \textbf{Key findings in PET quality} & \textbf{Author, year} \\ \hline 
U-net & Brain: 40, 5-fold cross validation & N/A* & Average 5\% error in activity quantification & Hwang \textit{et al.}, 2018\cite{RN5256} \\ \hline 
U-net & Brain: 100 training/28 testing & MAE (HU): 111$\mathrm{\pm}$16 & Bias: $\mathrm{<}$2\% among 28 VOIs & Liu \textit{et al.}, 2018\cite{RN5134} \\ \hline 
GAN & Brain: 50 training/40 testing & N/A & Bias: $\mathrm{<}$2.5\% among 7 VOIs & Armanious \textit{et al.}, 2018\cite{RN4765} \\ \hline 
U-net & Whole body: 60 training/20 validation/20 testing & Relative error (\%): 0.91$\mathrm{\pm}$3.55 (soft tissue)\newline 0.43$\mathrm{\pm}$6.80 (bone) & Bias (\%): 1.31$\mathrm{\pm}$3.55 in lesions & Hwang \textit{et al.}, 2019\cite{RN5471} \\ \hline 
CycleGAN & Whole body: 80 training/39 testing & MAE (HU): 108.9$\mathrm{\pm}$19.1 & Bias (\%): 1.07$\mathrm{\pm}$9.01 in lesions & Dong \textit{et al.}, 2019\cite{RN4754} \\ \hline 
\end{longtable}

*N/A: not available, i.e. not explicitly indicated in the publication

\bigbreak
\noindent 
\subsection{Intra-modality}

The group of intra-modality includes studies that transform between two different protocols within a same imaging modality, such as among different sequences of MRIs, or the restoration of images from low quality protocol to high quality protocol. Studies solely aiming at image quality improvement such as image denoising and artifact correction is not included in this study. Based on the studied image modalities, studies in this group can be further divided into CT, MR and PET groups. As shown in Figure 3, the number of studies on the three imaging modalities is close.  

\bigbreak
\noindent 
{\bf 4.2.1 CT}

CT imaging dose becomes an increasing public concern nowadays since excessive radiation dose can lead to increased risks of radiation-induced cancer and genetic defects.\cite{RN699, RN701, RN702} It is common for patients to undergo multiple CT scans in different procedures of diagnosis and treatment. Thus the accumulated imaging dose can be a big concern, particularly for pediatric patients who are more sensitive to radiation and have longer life expectancy than adults.\cite{RN1432}   
\bigbreak
       CT dose can be lowered by either reducing X-ray exposure (mAs).\cite{RN765, RN1431, RN722, RN1119} or the number of X-ray projections.\cite{RN765, RN1431, RN722, RN1119} However, if still reconstructing image by conventional filter-backprojection (FBP) algorithm, image quality would be degraded with elevated image noise level and reduced image signal-to-noise ratio for low mAs scheme, or with severe undersampling artifacts for low projection scheme. These low-quality images effects would make routine tasks on CT images difficult for clinicians. Hardware-based methods such as optimization on data acquisition protocol (automatic exposure control) \cite{RN703} and improvements in detector designs\cite{RN704} have been shown to be effective in reducing imaging dose to some extent while maintaining clinically acceptable image quality. However, further dose reduction from these techniques is limited by detector physical properties and hence very costly. 
\bigbreak
       Iterative CT image reconstruction algorithms have been proposed for decades to address the degraded image quality resulted from insufficient data acquisition.\cite{RN700} It models the physical process of a CT scan with prior knowledge, thus is more resilient to noise and require less imaging dose for the same image quality than FBP. \cite{RN706, RN200, RN700} However, iterative reconstruction suffers from long computing time due to the large number of iterations with repeated forward and back projection steps. Moreover, in the forward projection step, it requires the knowledge of energy spectrum which is difficult to directly measure.\cite{RN1434, RN1433, RN624, RN1435} It is usually simplified by a monoenergetic forward projection matrix, or having an indirect simulation/estimation of energy spectrum.\cite{RN1364, RN726, RN200, RN722, RN1119} 
\bigbreak
       Image synthesis by deep learning seems attractive for low dose CT restoration due to its data-driven approaches toward automatically learning image features and model parameters. As listed in Table VII, most of the methods in the reviewed literatures are direct image translation from low dose CT to full dose CT, while the others restore the sinogram using deep learning first, and then reconstruct images from the restored sonogram by FBP. As shown by Dong \textit{et al.}, with similar network, their proposed projection-based method outperformed an image-based method in better reducing downsampling artifacts with higher resolution on edges of the object.\cite{RN4738} A potential reason of such difference, as comment by the authors, is that for image-based method, the error in prediction is directly shown on image, while for projection-based method, the error predicted on sonogram will be compensated in the reconstruction process which is inherently a weighted sum. Thus projection-based method can be more tolerant to errors. It is also possible to train the model to map directly from projection domain to image domain, while the network must encode a change between polar and Cartesian coordinates.\cite{RN5472} Among image-domain methods, Shan et al used a progressive scheme that denoised the input low dose CT multiple times and yielded a sequence of denoised images at different noise levels.\cite{RN5473} In the study of Kang \textit{et al.}, instead of directly mapping low and full dose CT images, the proposed method mapped their wavelet coefficient. The benefit of wavelet transform was shown in better  structure recovery than direct image mapping.\cite{RN5474}  
\bigbreak
       Compared with iterative reconstruction methods, learning-based methods require less time and no prior knowledge about energy spectrum. For example, as reported by Wang \textit{et al.}, it takes about one minute to generate an entire 3D volume of their denoised low dose CT images on an average personal computer after their model is trained. In contrast, with the same hardware, a compressed-sensing-based iterative method needs one minute in forward and back projecting on a single slice in one iteration. Alternatively, if the forward and back projecting operation is pre-calculated and saved as a sparse matrix, the time shorten to several seconds for each slice in each iteration, but it requires 6.8 GB space in memory to store the matrix. Even so, to reconstruct the entire volume, it still needs several hours. Thus, it is very time and memory-consuming for conventional iterative reconstruction method to be implemented on personal computer, especially when slices thickness is small and FOV on slice-direction is large.\cite{RN4740} 
\bigbreak
       Conventional iterative reconstruction methods were compared with learning-based methods in several studies. For example, total variation (TV) regularization is commonly studied in state-of-the-art compressed sensing-based iterative methods. A common finding is that TV-based methods tend to over-smooth and present patchy texture, while the results by learning-based methods have finer structures preserved and closer to a full dose CT in image texture. \cite{RN5472, RN4740} Such improvement of learning-based methods is also shown in quantitative metrics of PSNR, etc. The large recovery of image quality with a preserved image texture could be attributed to the learning process that the prediction images always tend to be trained to have a similar image quality and texture as its target images, i.e. full dose CT images. Similarly, Shan \textit{et al.} demonstrated that their proposed learning-based method performed favorably or comparably to three commercially available iterative algorithms in terms of noise suppression and structural fidelity by double-blinded reader study.  
\bigbreak
       Most of the reviewed studies assume the application of their restored full dose CT to be diagnosis. Wang \textit{et al.} evaluated their method in the context of radiation therapy treatment planning.\cite{RN4740} Their motivation is that low dose CT in CT simulation process is attractive for adaptive radiation therapy where multiple re-scanning and re-planning during treatment course is very common. In contrast to the diagnosis CT that pursues high spatial resolution and low-contrast detectability, planning CT requires accurate HU numbers and dose calculation accuracy. Their dosimetry study showed that the average differences of DVH metrics between the synthetic full dose CT and original full dose CT are less than 0.1 Gy (p$>$0.05) when prescribed dose is 21 Gy. 
\bigbreak
       Many of the reviewed studies used the dataset from the AAPM 2016 Low-dose CT Grand Challenge.\cite{RN5487} Although the training and testing scheme may be different among these studies, the results of these studies can still have a fair comparison. However, this low dose CT dataset, along with the dataset in most of the other studies, is simulated from full dose CT by adding Poisson noise or downsampling sinogram, due to the lack of clinical low dose CT data. Exceptions are Yi \textit{et al.} using piglet and Shan \textit{et al.} using real patient low dose CTs.\cite{RN5473, RN5478} The simulated noise may not fully reflect the noise level and potential artifacts, thus it is of clinical interest to evaluate these methods with physically measured low dose datasets. 

\bigbreak
\noindent \textbf{Table VII.} Summary of studies on synthetic full dose CT from low dose CT.

\begin{longtable}{|p{0.7in}|p{0.8in}|p{0.9in}|p{0.9in}|p{0.8in}|p{0.9in}|} \hline 
\textbf{Network} & \textbf{Projection or image domain} & \textbf{Site, and \# of patients in training/testing} & \textbf{Low dose scheme and fraction of full dose CT} & \textbf{Key findings in restored full dose CT} & \textbf{Author, year} \\ \hline 
U-net & Image & Abdomen: 10 training/20 testing & Low mAs: 1/4 of full dose & PSNR (dB): about 36 & Kang \textit{et al.}, 2017\cite{RN5474} \\ \hline 
U-net & Image & Abdomen: 10, leave-one-out & Low mAs: 1/4 of full dose & PSNR (dB): 44.4187$\mathrm{\pm}$1.2118 & Chen \textit{et al.}, 2017\cite{RN5012} \\ \hline 
U-net & Image & Thorax and pelvis: 475 slices training/25 slices testing & Sparse view: 1/20 of full views & PSNR (dB): 28.83 & Jin \textit{et al.}, 2017\cite{RN5472} \\ \hline 
GAN & Image & Cardiac: 28, 2-fold cross validation & Low mAs: 20\% dose & Significantly reduced noise & Wolterink \textit{et al.}, 2017\cite{RN5476} \\ \hline 
CNN(ResNet) & Image & Abdomen: 9 training/1 testing & Low mAs: 1/4 of full dose & PSNR (dB): 39.8329 & Yang \textit{et al.}, 2017\cite{RN5480} \\ \hline 
CNN (ResNet) & Image & Abdomen: 8 training/1 testing & Low mAs: 1/4 of full dose & PSNR  (dB): 38.70 & Kang \textit{et al.}, 2018\cite{RN5477} \\ \hline 
GAN & Image & Abdomen (piglet): 708 slices training/142 slices testing & Low mAs: 5\% of full mAs & PSNR (dB): about 34 & Yi \textit{et al.}, 2018\cite{RN5478} \\ \hline 
GAN & Image & Abdomen: 5 training/5 testing & Low mAs: 1/4 of full dose & PSNR(dB): 30.137$\mathrm{\pm}$1.938 & Shan \textit{et al.}, 2018\cite{RN5479} \\ \hline 
GAN & Image & Abdomen: 10, leave-one-out & Low mAs: 1/4 of full dose & PSNR (dB): (25.372, 27.398)${}^{\ +}$ & You \textit{et al.}, 2018\cite{RN5482} \\ \hline 
U-net & Image & Abdomen: 8 training/1 validation/1 testing & Sparse view: 1/12 of full views & PSNR (dB): 40.4856 & Han \textit{et al.}, 2018\cite{RN5484} \\ \hline 
GAN & Image & Abdomen: 4000 slices training/2000 testing & Low mAs: 1/4 of full dose & Validated in double-blinded reader study & Yang \textit{et al.}, 2018\cite{RN5475} \\ \hline 
U-net (Encoder-decoder) & Image & Whole body: 300 slices training/50 slices testing & Low mAs: fraction not specified & PSNR (dB): 42.3257 & Liu \textit{et al.}, 2018\cite{RN5486} \\ \hline 
CNN & Image & Chest: 3 training/3 testing & Low mAs: 3\% of full mAs & PSNR (dB): about 22 & Zhao \textit{et al.}, 2019\cite{RN4878} \\ \hline 
U-net & Projection & Chest: 7 training/8 testing & Sparse view: 1/4 of full views & PSNR (dB): (42.73, 52.14) & Lee \textit{et al.}, 2019\cite{RN5481} \\ \hline 
U-net & Image & Abdomen and chest: 10 training/60 testing & Low mAs: about 1/3 to 1/8 of full dose & Validated in double-blinded reader study & Shan \textit{et al.}, 2019\cite{RN5473} \\ \hline 
U-net & Projection & Head: 200 slices training/100 slices testing & Sparse view: 1/12 of full views\newline Limited angle: 1/4 of full views & PSNR (dB): 37.21 for sparse view\newline 43.69 for limited angle & Dong \textit{et al.}, 2019\cite{RN4738} \\ \hline 
CycleGAN & Image & Head: 30, 5-fold cross validation & Low mAs: 0.5\% of full mAs & NMSE (\%): 1.63$\mathrm{\pm}$0.62 & Wang \textit{et al.}, 2019\cite{RN4740} \\ \hline 
\end{longtable}

${}^{+}$Numbers in parentheses indicate minimum and maximum values.

\bigbreak
\noindent 
{\bf 4.2.2 MRI}

Image synthesis has been investigated for various applications in MR imaging,\cite{RN5515} such as imaging translation among different sequences, converting low-magnetic-field MRI to high-magnetic-field MRI, and restoring undersampled acquisition. The motivation of converting a low-magnetic-field MRI to high-magnetic-field MRI is to acquire MR images by an accessible MRI scanner but with high spatial resolution and good contrast as on a cutting-edge scanner. The translation among different sequences and restoring undersampled acquisition can both shorten the acquisition time. The challenges in the former are mapping the different contrast, and in the latter are recovering the spatial resolution. Although these applications are not exactly the same in terms, they pose similar problems for image synthesis, i.e. preserving contrast and resolution.  
\bigbreak
       A large group of conventional methods that address these problems are compressed sensing methods assuming that images have a sparse representation in some transform domain. For example, in image synthesis among multi-contrast MRI, the image patch of one contrast is expressed as a sparse linear combination of patches in atlas, and such combination is then applied on the image patches on the other contrast. In recovering undersampled acquisition, the problem is usually modeled as a reconstruction problem with regularization terms that incorporate prior knowledge about the sparsity of images. It is usually implemented as an iterative algorithm to solve the optimization problem, which is time and resource intensive. Another property of CS is that it requires incoherence in the sampling scheme, which can be disadvantageous in some cases such as when downsampling in k-space region is preferable.\cite{RN5228} On the other hand, the success of deep learning in other imaging synthesis fields encourages the integration of neural network into the above situations for its favorable performance in short prediction time as well as non-linear mapping capability. 
\bigbreak
       The related studies are listed in Table VIII. Compared with image synthesis in other applications, more reviewed studies of MR inter-modality incorporate neural network with other techniques, rather than a direct end-to-end method. It is also common to apply neural network in transform domains. Zhang \textit{et al.} proposed a cascaded regression using two parallel and interactive multi-layer network streams on spatial and frequency domains. Compared with using single spatial-domain, the dual-domain method presented better visual results and significantly larger SSIM.\cite{RN4894} Qu \textit{et al.} designed a wavelet-based affine transformation layer to modulate feature maps from spatial domain and wavelet domain in the encoder, followed by the image reconstruction decoder branch that synthesizes 7T images from the wavelet modulated spatial information. Without such layer, the framework was reduced to a plain encoder-decoder network, which was found to be less capable in recovering details.\cite{RN5488} 
\bigbreak
       Many of the reviewed studies also compared their proposed method with CS-based method. With a comparable or better performance on quantitative image quality metrics, a noticeable improvement is much less computational time. The predication of learning-based method is usually in a magnitude of millisecond to second, while CS-based method is in minutes. Schlemper \textit{et al.} found that at low undersampling rate, learning-based and CS-based methods had comparable performance, and the advantageous performance of learning-based method was shown at more aggressive undersampling factor.\cite{RN4997} Other findings of blurred details and blocky artifacts as mentioned in previous sections were also reported. 
\bigbreak
\noindent \textbf{Table VIII.} Summary of studies on synthetic MRI.

\begin{longtable}{|p{0.5in}|p{1.2in}|p{1.2in}|p{1.2in}|p{0.9in}|} \hline 
\textbf{Network} & \textbf{Applications} & \textbf{Site, and \# of patients in training/testing} & \textbf{Key findings in results} & \textbf{Author, year} \\ \hline 
GAN & Synthesizing 7T MRI from 3T MRI & Brain: 15, leave-one-out & PSNR (dB): 27.6$\mathrm{\pm}$1.3 & Nie \textit{et al.}, 2018\cite{RN5196} \\ \hline 
CNN & Restoring undersampled acquisition & Cardiac: 5 training/5 testing & Restored images showed most of the anatomical structures up to 11-fold undersampling & Schlemper \textit{et al.}, 2018\cite{RN4997} \\ \hline 
GAN & Low resolution to high resolution & Brain: 196 training/48 testing & SSIM: (0.76, 0.94)${}^{\ +}$ at 8-fold undersampling & Kim \textit{et al.}, 2018\cite{RN5228} \\ \hline 
U-net & Synthesizing full contrast enhanced images from low contrast enhanced images & Brain: 10 training/ 50 testing & PSNR (dB): 28.07$\mathrm{\pm}$2.26 at 10-fold lower  & Gong \textit{et al.}, 2018\cite{RN5257} \\ \hline 
U-net & T1w to T2w \newline T1w to FLAIR\newline PDw to T2w & Three different brain datasets: 22 training/3 validation/3 testing\newline 42 training/6 validation/6 testing\newline 22 training/3 validation/3 testing\newline  & Average PSNR (dB) among groups of datasets: (25.78, 32.92) for synthetic T2w\newline (29.99, 30.32) for synthetic FLAIR & Chartsias \textit{et al.}, 2018\cite{RN5495} \\ \hline 
GAN & Restoring undersampled acquisition & Brain and chest: for each site, 100 slices training/100 slices testing & PSNR (dB) at 10\% undersampling: about 32 for brain, 26.5 for chest & Quan \textit{et al.}, 2018\cite{RN5491} \\ \hline 
\multirow{5}{*}{GAN} & Low resolution to high resolution & 767 training/192 validation/30 testing & \multirow{5}{*}{\makecell{Average PSNR (dB):\\ about (25, 30)}}  & \multirow{5}{*}{\makecell{Galbusera \textit{et al.},\\ 2018\cite{RN5141}}} \\ 
 & T1w to T2w & 767 training/192 validation/30 testing &  &  \\ 
 & T2w to T1w & 767 training/192 validation/30 testing &  &  \\ 
 & T2w to STIR & 284 training/71 validation/30 testing &  &  \\ 
 & T2w to TIRM & 305 training/77 validation/30 testing &  &  \\ \hline 
GAN & T1w to T2w\newline T2w to T1w & Brain: 3 datasets, 48 training/5 validation/13 testing,\newline 25 training/5 validation/10 testing,\newline 24 training/2 validation/15 testing & Average PSNR (dB): (25.80$\mathrm{\pm}$1.87, 29.77$\mathrm{\pm}$1.57) among three datasets & Dar \textit{et al.}, 2019\cite{RN5494} \\ \hline 
GAN & Restoring undersampled acquisition & Abdomen: 336 training/10 testing & SSIM: 0.84 at 5-fold undersampling & Mardani \textit{et al.}, 2019\cite{RN5492} \\ \hline 
U-net & Synthesizing DTI from fMRI & Brain: 648 training/293 testing & Mean correlation coefficient: 0.808$\mathrm{\pm}$0.054 among 38 VOIs & Son \textit{et al.}, 2019\cite{RN4722} \\ \hline 
CNN & Synthesizing FLAIR from mpMRI & Brain: 24, 5-fold cross validation & SSIM: 0.860$\mathrm{\pm}$0.031 & Wei \textit{et al.}, 2019\cite{RN5067} \\ \hline 
GAN & Synthesizing diffusion b0 maps from T1w & Brain: 586 training/26 testing & Distortion correction based on synthesized b0 maps is feasible & Schilling \textit{et al.}, 2019\cite{RN4924} \\ \hline 
CNN (ResNet) & Synthesizing arterial spin labeling images from T1w & Brain: 355, 5-fold cross validation & Accuracy in CBF calculation and dementia disease diagnosis is close to gold standard & Huang \textit{et al.}, 2019\cite{RN4962} \\ \hline 
CNN & Synthesizing 7T MRI from 3T MRI & Brain: 15, leave-one-out & SSIM: 0.8438 & Zhang \textit{et al.}, 2019\cite{RN4894} \\ \hline 
U-net & Restoring undersampled acquisition & Knee: 90 training/10 validation/10 testing & SSIM: 0.821$\mathrm{\pm}$0.023 at 8-fold undersampling & Liu \textit{et al.}, 2019\cite{RN4975} \\ \hline 
U-net (encoder-decoder) & Synthesizing 7T MRI from 3T MRI & Brain: 15, leave-one-out & PSNR (dB): 28.27 & Qu \textit{et al.}, 2020\cite{RN5488} \\ \hline 
U-net & Restoring undersampled acquisition & Knee: 336 training/24 testing & SSIM: 0.8603 at 4-fold undersampling & Wu \textit{et al.}, 2020\cite{RN4971} \\ \hline 
U-net & Synthesizing MR angiography from 3D-QALAS sequence & Brain: 11, 5-fold cross validation & PSNR (dB): 35.3$\mathrm{\pm}$0.5 & Fujita \textit{et al.}, 2020\cite{RN4628} \\ \hline 
\end{longtable}

${}^{+}$Numbers in parentheses indicate minimum and maximum values.

\bigbreak
\noindent 
{\bf 4.2.3 PET}

Image synthesis among different PET images has been proposed to facilitate PET AC and low-count PET reconstruction. For the PET AC, unlike what is mentioned in section 4.1.4 that synthetic CT is generated from NAC PET for attenuation correction during PET reconstruction, a few studies, as listed in Table IX, investigated the feasibility of direct mapping NAC PET to AC PET by exploiting the deep learning methods to bypass the PET reconstruction step. These studies reported comparable results with synthetic CT-based PET AC (section 4.1.4), while a direct comparison on same datasets were not found. Dong \textit{et al.} applied the direct NAC PET-AC PET mapping on whole body for the first time.\cite{RN4668} They also demonstrate the reliability of their method by including sequential scans in their testing datasets to evaluate the PET intensity changes with time on their AC PET as well as ground truth. Similar as their study using synthetic CT, lung showed the largest error. Shiri \textit{et al.} further assessed the radiomic features on their AC PET results, and found only 3 out of 83 regions had significant difference with ground truth.\cite{RN4874} 
\bigbreak
       Low-count PET has extensive application in pediatric PET scan and radiotherapy response evaluation with advantage of better motion control and low patient dose. However, the low count statistics would result in increased image noise, reduced contrast-to-noise ratio, and large bias in uptake measurement. The reconstruction of a standard- or full-count PET from the low-count PET cannot be achieved by simple postprocessing operations such as denoising since lowering radiation dose changes the underlying biological and metabolic process, leading to not only noise but also local uptake values changes.\cite{RN1837} Moreover, even with a same tracer injection dose, the uptake distribution and signal level can vary greatly among patients. The learning-based low-count PET reconstruction methods, as summarized in Table X, have been proposed to take advantage of their powerful data-driven feature extraction capabilities between two image datasets. A few of the reviewed methods used both MR and low-count PET as input, while most used low-count PET only. Most proposed methods were implemented on PET brain scans, with a few on lung and whole body. Compared with the evaluations in PET AC which focus on relative bias, the evaluations in the reviewed studies in low-count PET reconstruction more focus on the image quality and the similarity between the predicted result and its corresponding full-count PET ground truth.  
\bigbreak
       Similar as low dose CT restoration, most low-count PET restoration studies applied neural network directly on image domain, with a few on projection domain. In addition to the advantages mentioned in section 4.2.1, Sanaat \textit{et al.} commented that projection-based network allows the change of reconstruction filter or any post-processing without the need of retrain the model.\cite{RN4647} They also compared the results by the same proposed network using image and projection as input, and found that projection-based results better reflected uptake pattern and anatomy than image-based results, and both subjective and objective studies validated the advantages of projection-based results. A drawback of projection-based method is six times longer training time than image-based method. 
\bigbreak
       Although these studies demonstrate the feasibility of mapping low-count PET to full-count PET, a few studies investigated using both PET and MRI image as dual input channels to further improve the results when MR images are available. As expected, the addition of MRIs that provides anatomical information could help improve the performance of the network than without MR. Chen \textit{et al.} showed that their network is able to achieve 83\% accuracy when using only PET as input, and 89\% when using PET+MR, in a clinical reading study of uptake status.\cite{RN5107} The potential reason of such difference lies in that the results by PET+MR were superior in reflecting the underlying anatomic patterns. The contribution of MR images was also validated in the study of Xiang et al by a significant improved PSNR.\cite{RN4988} They commented that structural information from MRIs yields important cues for estimating the high-quality PET, even though structural MRIs differ from PETs significantly regarding their appearances. 

\bigbreak
\noindent \textbf{Table IX.} Summary of studies on synthetic AC PET from NAC PET.

\begin{longtable}{|p{0.8in}|p{1.4in}|p{1.4in}|p{1.4in}|} \hline 
\textbf{Network} & \textbf{Site, and \# of patients in training/testing} & \textbf{Key findings in PET quality} & \textbf{Author, year} \\ \hline 
U-net & Brain: 25 training/10 testing & Bias (\%): 4.0$\mathrm{\pm}$15.4 & Yang \textit{et al.}, 2019\cite{RN1828} \\ \hline 
U-net & Brain: 91 training/18 testing & Bias (\%): -0.10$\mathrm{\pm}$2.14 among 83 VOIs & Shiri \textit{et al.}, 2019\cite{RN4874} \\ \hline 
CycleGAN & Whole body: 25 training/\newline  10 patients*3 sequential scan testing & Bias (\%): (-17.02,3.02)${}^{\ +}$ among 6 VOIs, 2.85$\mathrm{\pm}$5.21 in lesions & Dong \textit{et al.}, 2019\cite{RN4668} \\ \hline 
\end{longtable}

${}^{+}$Numbers in parentheses indicate minimum and maximum values.

\bigbreak
\noindent \textbf{Table X.} Summary of studies on synthetic full count PET from low count PET.

\begin{longtable}{|p{0.5in}|p{0.7in}|p{0.7in}|p{0.9in}|p{0.7in}|p{0.7in}|p{0.8in}|} \hline 
\textbf{Network} & \textbf{PET or PET+MR} & \textbf{Image or projection domain} & \textbf{Site, and \# of patients in training/testing} & \textbf{Counting fraction (low/full)} & \textbf{Key findings in restored full counting PET} & \textbf{Author, year} \\ \hline 
CNN  & PET+MR & Image & Brain: 16, leave-one-out & 1/4 & PSNR (dB): 24.76  & Xiang \textit{et al.}, 2017\cite{RN4988} \\ \hline 
GAN & PET & Image & Brain: 16, leave-one-out & 1/4 & PSNR (dB): about 24 & Wang \textit{et al.}, 2018\cite{RN1842} \\ \hline 
U-net  & PET+MR & Image & Brain: 40, five-fold cross-validation & 1/100 & PSNR (dB): about 38 & Chen \textit{et al.}, 2019\cite{RN5107} \\ \hline 
CNN & PET & Image & Brain: 2 training/1 testing\newline Lung: 5 training/1 testing & Brain: 1/5\newline Lung: 1/10 & N/A & Gong \textit{et al.}, 2019\cite{RN1860} \\ \hline 
U-net (encoder-decoder) & PET & Projection & Whole body (simulation): 245 training/52 validation/53 testing & N/A & PSNR (dB): 34.69 & H\"{a}ggstr\"{o}m \textit{et al.}, 2019\cite{RN4952} \\ \hline 
GAN  & PET+MR & Image & Brain: 16, leave-one-out & ? & PSNR (dB): 24.61  & Wang \textit{et al.}, 2019\cite{RN1843} \\ \hline 
GAN & PET & Image & Brain: 40, four-fold cross validation & 1/100 & PSNR (dB): about 30 & Ouyang \textit{et al.}, 2019\cite{RN4908} \\ \hline 
U-net & PET & Image & Lung: 10, five-fold cross validation & 1/10 & Bias: $\mathrm{<}$15\%  & Lu \textit{et al.}, 2019\cite{RN4845} \\ \hline 
GAN & PET & Image & Whole body: 435 slices training/440 slices testing & 1/10 & PSNR (dB): 30.557 & Kaplan \textit{et al.}, 2019\cite{RN5483} \\ \hline 
CycleGAN & PET & Image & Whole body: 25 training/10 testing & 1/8 & PSNR (dB): 41.5$\mathrm{\pm}$2.5 & Lei \textit{et al.}, 2019\cite{RN4770} \\ \hline 
U-net  & PET & Projection & Brain: 100 training/20 validation/20 testing & 1/20 & PSNR (dB): 38.25$\mathrm{\pm}$0.66 & Sanaat \textit{et al.}, 2020\cite{RN4647} \\ \hline 
\end{longtable}

\bigbreak
\noindent 
\section{ SUMMARY AND OUTLOOK}

Recent years have witnessed the trend of deep learning being increasingly used in the application of medical imaging. The latest networks and techniques have been borrowed from computer vision field and adapted to specific clinical tasks for radiology and radiation oncology. As reviewed in this paper, learning-based image synthesis is an emerging and active field since all of these reviewed studies were published within the recent three years. With the development in both artificial intelligence and computing hardware, more learning-based methods are expected to facilitate the clinical workflow with novel applications. Although the reviewed literatures show the success of deep learning-based image synthesis in various applications, there are still some common open questions that need to be answered in future studies.  
\bigbreak
       In the implementation of the network, due to the imitations of the GPU memory, some of the deep learning approaches are trained on two-dimensional (2D) slices. Since the loss functions of 2D models do not account for continuity in the third dimension, slice discontinuities can be observed. Some studies trained models in 3D patches to exploit 3D information with even less memory burden,\cite{RN4913} while a potential drawback is that the larger scale image features may be hard to extract.\cite{RN4885} Training on three-dimensional (3D) image stacks is expected to achieve a more homogeneous conversion result. Fu \textit{et al.} compared the performance between 2D and 3D model using the same U-net.\cite{RN4879} They found that 3D model generated synthetic CT with smaller MAE and more accurate bone region. However, to achieve robust performance, 3D model needs more training data since it has more parameters. A compromised solution is to use multiple adjacent slices that may allow the model to capture more image context information, or to train different networks for all three orthogonal 2D planes to allow pseudo 3D information.\cite{RN4924} 
\bigbreak
        The reviewed studies showed the advantages of learning-based methods over conventional methods in performance as well as clinical applications. Learning-based methods generally outperform conventional methods in generating more realistic synthetic images with higher similarity to real images and better quantitative metrics. In implementation, depending on the hardware, training a model usually takes several hours to days for learning-based methods. However, once the model is trained, it can be applied to new patients to generate synthetic images within a few seconds or minutes. Conventional methods vary a lot in specific methodologies and implementations, resulting in a wide range of run time. Iterative methods such as CS were shown to be unfavorable for large time and resource consuming. 
\bigbreak
       In the training stage, most of the reviewed studies require paired datasets, i.e. the source image and target image need to have pixel-to-pixel correspondence. This requirement poses difficulties in collecting sufficient eligible datasets, as well as demands high accuracy in image registration. Some networks such as CycleGAN can relax the requirement of the paired datasets to be unpaired datasets, which can be beneficial for clinical application in enrolling large number of patient datasets for training.  
\bigbreak
       Although the advantages of learning-based methods have been demonstrated, it should be noted that its performance can be unpredictable when the input images are very different from its training datasets. In most of the reviewed studies, unusual cases are excluded. However, these unusual cases can happen from time to time in clinic, and should be dealt with caution. For example, it is not uncommon to see patient with hip prosthesis in pelvis scan. The hip prosthesis creates severe artifacts on both CT and MR images, thus, it can be of clinical interest to see the related effect of its inclusion in training or testing dataset, which has not been studied yet. Similar unusual cases can also be seen in other forms in all imaging modalities and are worth investigation, including all kinds of implants that can introduce artifacts, obese patients that present much higher noise level on image than average, and patients with anatomical abnormality. 
\bigbreak
       Due to the limitation in the number of available datasets, most studies used N-fold cross validation or leave-N-out strategy. The small to intermediate number of patients in training/testing is proper for feasibility study, while is far from enough in evaluating clinical utility and potential impact. Moreover, the representativeness of training/testing dataset needs special attention in clinical study. The missing of diverse demographics may reduce the robustness and generality in the performance of the model. Most of the studies trained model using data from a single institution with a single scanner. As replacing/equipping with new scanner is common in practice, it is interesting to know how the trained model would perform on another scanner of different model or vendor when the image characteristic cannot be exactly matched. Boni \textit{et al.} recently presented a proof-of-concept study that predicted synthetic images of one site using model trained on another two sites, and demonstrated the clinical acceptable synthetic results.\cite{RN5497} Further studies could include datasets from multiple centers and adopted a leave-one-center-out training/test strategy in order to validate the consistency and robustness of the network. 
\bigbreak
       Before being deployed into clinical workflow, there are still a few challenges to be addressed.  To account for the potential unpredictable synthetic images that can be resulted by noncompliance with imaging protocols as training data, or unexpected anatomic structures, additional quality assurance (QA) step would be essential in clinical practice. The QA procedure would aim to check the consistency on the performance of the model routinely or after upgrade by re-training the network with more patient datasets, as well as to check the synthetic image quality of patient-specific case. 

\bigbreak
\noindent 
\section{ ACKNOWLEDGEMENT}

This research was supported in part by the National Cancer Institute of the National Institutes of Health under Award Number R01CA215718 and Emory Winship Cancer Institute pilot grant.

\noindent 

\bibliographystyle{plainnat}  
\bibliography{Manuscript_tex1}       

\end{document}